\documentclass{article}
\RequirePackage{etex}
\usepackage{graphicx} 
\usepackage[hmargin=1in,vmargin={1.3in,1.3in}]{geometry}
\usepackage{amsthm}
\usepackage{amsmath}
\usepackage{amsfonts}
\usepackage{amssymb}
\usepackage{titling} 
\usepackage[affil-it]{authblk}

\usepackage{csquotes}
\usepackage{kpfonts}
\usepackage{color}
\usepackage{amsmath, amssymb, enumitem, verbatim, color, tikzsymbols, tabularx, dcolumn,longtable, array, pdflscape, soul, bbm, subcaption, rotating}

\usepackage[titletoc,toc,title]{appendix}
\usepackage{verbatimbox} 
\usepackage{etoolbox}
\usepackage{environ}
\NewEnviron{myresizeenv}{\resizebox{\linewidth}{!}{\BODY}}

\usepackage{tocloft}
\makeatletter
\renewcommand{\numberline}[1]{{\@cftbsnum #1\@cftasnum~}\@cftasnumb}
\makeatother

\usepackage{booktabs}
\usepackage{array}
\usepackage{siunitx}
\usepackage{multirow, multicol}
\usepackage{textcomp}
\usepackage{float}

\usepackage{subcaption, caption} 
\usepackage[style=authoryear, citetracker=true, maxcitenames=2, hyperref=true, url=true, isbn=false, doi=true, natbib=true, backend=biber,bibstyle=authoryear, maxbibnames=30, sorting=nyt, uniquename=full, uniquelist=false]{biblatex}
\usepackage[colorlinks=true, allcolors=blue, hyperindex,breaklinks]{hyperref}
\usepackage{setspace}
\usepackage{placeins, adjustbox}
\usepackage{ragged2e}

\title{
Mapping the disaggregated economy in real-time: \\Using granular payment network data to complement national accounts 
}
\author[a]{Kerstin Hötte\footnote{Corresponding author: kerstin.hotte@kedgebs.com \\An earlier version of this article circulated under the title \emph{National accounting from the bottom up using large-scale financial transactions data: An application to input-output tables}.}}
\tiny 
\affil[a]{KEDGE Business School, Paris}
\date{\today}
\begin{document}

\maketitle
\begin{abstract}
     In an era of rapid change, timely and disaggregated economic insights are crucial for effective policymaking. This study explores the potential of real-time payment data to complement traditional economic measurement. Using anonymised UK business payments from 2015–2023, we analysed inter-industry financial flows at a granular 5-digit SIC level and compared them systematically with established economic indicators such as GDP and input-output tables (IOTs). Our findings show strong correlations with GDP and qualitative consistency with official IOTs, highlighting the value of the novel high-frequency data for real-time economic monitoring. We also benchmarked network statistics at the 5-digit level, showing how industry-specific payment structures align with stylised facts from the empirical economic network literature. While outlining methodological and interpretative challenges, we discuss the integration of such bottom-up data into national accounts. This work contributes to ongoing efforts to advance economic measurement and offers additional tools for tracking economic dynamics in real time. 
     
\end{abstract}
\vspace{0.5cm} \noindent

\noindent
\vspace{0.2cm}

\newpage
\section*{Acknowledgments}
The author acknowledges valuable contributions and feedback by Andreina Naddeo, whose work helped to develop and improve this research. Further gratitude is owed to Johannes Lumma for his research efforts and contributions to an earlier version of this article, and to Fran\c{c}ois Lafond for his contributions to the design of the study and management of the project. 
The author would like to thank her colleagues at the ONS and Vocalink, especially Dragos Cozma, Nathan Williams, Mark Greenberg and Joseph Colliass for technical advice and data development, and Keith Lai and Victor Meirinhos for making the data available for research. 
This work has also benefited from the support of colleagues at the Alan Turing Institute and the Universities of Oxford and Cambridge. 
\section*{Disclaimer}
This work has been produced within a partnership between the Office for National Statistics and the Alan Turing Institute to explore the development of innovative payment systems data to provide new economic insights and improve official economic statistics. 
The views expressed are those of the author and do not reflect those of the Office for National Statistics or the wider UK Government. 

\newpage

\section{Introduction}
The grand policy challenges of today require a granular understanding of our economy, ideally in real time. 
Examples include supply chain disruptions caused by pandemics, climatic shocks and regional conflicts, or transition policies for net-zero and economic resilience. 
New and large-scale data can help tackle these challenges, and statistical offices are currently exploring how such data can be developed and used at a macroeconomic scale \citep{UN2023, ons2023realtime, whitehouse2023supply, woloszko2023nowcasting, swetkis2013putting, he2024electricity}. 
New data come with new challenges, and it is unclear how to interpret the data in the established terminology of national accounts (NAs), developed almost a century ago \citep{kuznets1937national}. NAs adopt an aggregate view on markets while hiding the realised patterns of exchange between the trading firms \citep{simon1995organizations}. Data based on realised payments and aggregated from the bottom up can complement the top-down data when studying our macroeconomy, with a large potential for novel applications \citep{mantziou2023gnar, mantziou2024gdp}. 

In this paper, we use monthly experimental data on inter-industrial payments compiled from anonymised aggregates extracted from the Bankers' Automated Clearing Services (Bacs) payments system and provided to the Office for National Statistics (ONS). Bacs system is one of the major payment systems businesses use in the UK \citep{ons2023interindustry, payuk2013bacsprinciples}. 
These data are embedded in a series of other real-time indicators explored by the ONS \citep{ons2023realtime, ons2022revolut} and offer an unprecedented view of the UK economy and its supply chains. 
The data include a monthly network time series of industry-to-industry payments at 5-digit Standard Industry Classification (SIC) level and cover the period from August 2015 to December 2023, and bear the potential to be sourced in real time. 
Such granular real-time data on industry-to-industry flows had never been available before, while official inter-industry input-output tables (IOTs) are much more aggregate, published with time lags and only available at an annual frequency. 

However, indicators developed using inter-industry payments are new, and their usefulness in real-world economic analyses remains to be proven. Also, the interpretation of observed time trends, short-term responses to shocks, and static properties and their relationship to established economic indicators is not necessarily clear. For example, trends in payment aggregates can reflect actual economic dynamics or changes in payment preferences and behaviour. 
For example, cash transactions declined over the past decade, accelerated by Covid-19 \citep{UKfinance2022UKpayments}, while cashless payments have steadily increased \citep[][]{bodley2022fintech}. Such behavioural changes are independent of the underlying trends in ``real'' economic activity. 

Other challenges are posed by financial intermediation, which may respond to innovation and regulation in payment systems. Financial intermediaries execute transactions on behalf of their clients. Intermediation activities may inform about financial liquidity in the real economy (which is valuable information), but they hide the actual production activities and input-output links between trading industries.

Despite these and other challenges, the new data offer an unprecedented potential to advance research: beyond the timeliness and lower aggregation, our payment data offer entirely new data types, such as differentiations between the counts and values of transactions, entailing distinct kinds of economic information. 
On the downside, the inter-industrial payment data do not reveal the whole picture: depending on the purpose and type of transaction, businesses rely on multiple payment systems next to Bacs, such as card payments, high-value-high-security or international systems. 

This paper guides how to read the novel data and offers a validation exercise, showing how real-time payments relate to official macroeconomic time series and IOTs published by the \citet{ons2023blue}. A one-by-one validation is not possible in all dimensions, as monthly or 5-digit IOTs do not exist. Therefore, we rely on monthly macroeconomic indicators, annual IOTs, and stylised facts of granular production networks. 

We found that transaction values show strong statistical relationships to nominal economic indicators, while counts (the number of monthly transactions) appear powerful in picking up trends of data in real terms. 
To date, count data has rarely been used in economics, but it can be indicative of business dynamism: variations in the counts can reflect deviations from standing regular fixed and variable costs, including baseline intermediate purchases, fees, royalties, and loan repayments. 
We observe high auto-correlations and promising cross-correlations when comparing our inter-industrial payments to official IOTs and GDP. 
We supplement our quantitative analysis with a conceptual discussion of major sources of observed differences. These are, for example, the treatment of investments in physical capital, the financial and retail sector, and international trade, along with aspects related to classification and the time of recording. 

We also found that the structure of the highly granular 5-digit SIC payment network matches relevant stylised facts from the literature, such as correlating growth rates among neighbouring industries and centrality distributions \citep{carvalho2014from, mungo2023revealing, magerman2016heterogeneous, bacilieri2022firm}. This paves the way for applied economic research exploiting the granular network structure \citep{mantziou2024gdp}. This is a promising endeavour, as a long-time series of an evolving monthly proxy IOT at such a granular level has never been available to economic research before (to the best of our knowledge). 

This work relates to two major streams of research and data advances. Firstly, we contribute to recent and ongoing work on real-time but non-standardised data, fuelled by data science and new technology in economic measurement \citep{ons2023realtime, ons2022revolut, BoE2023technology, ialongo2022reconstructing, woloszko2023nowcasting, dietzenbacher2013input}. Our research provides an in-depth analysis of research challenges and the relations to official NAs, which may be similar to other bottom-up collected data on production networks. 
Secondly, our work embeds in current economic research trends on highly granular production network data, including networks reconstructed from financial transaction data \citep{fujiwara2021money, silva2022modeling, barja2019assessing, magerman2016heterogeneous}. We contribute by assessing data extracted from the payment system infrastructure, which adds a novel granular time-series dimension. 

The structure of this article is as follows: we provide an introduction to the UK payment systems in Sec. \ref{sec:payments_explained}. In \ref{sec:macro_benchmarking}, we assess the data at the macroeconomic level, before diving into the benchmarking against NAs (Sec. \ref{sec:benchmarking}). Sec. \ref{sec:conceptual} discusses the conceptual differences to NAs. Sec. \ref{sec:stylised_facts_5digit} explores stylised facts of the granular network. Sec. \ref{sec:conclusion} concludes.

\section{Data}
\label{sec:payments_explained}
Here we explain the main payment routes in the UK, focusing on business-to-business (B2B) transactions. This helps to understand the potential and limitations of payments as a data source. 
In the appendix, we provide additional detail on the basic concepts of payment systems (\ref{app:subsec:paysystems_basics}) and a short discussion of the possible impact of regulation and innovations on payment data (\ref{app:subsec:innovation_payments_explained}). 
\label{subsec:UK_payments_explained}

\begin{table}

    \caption{Overview of major UK payment schemes}
    \label{tab:overview_UKschemes}
    \centering \scriptsize
    \begin{tabular}{@{\extracolsep{2pt}} p{0.08\textwidth}p{0.08\textwidth}p{0.3\textwidth}p{0.3\textwidth}p{0.1\textwidth}}    
\\[-1.8ex]\hline 
\hline \\[-1.8ex] 
Name& Type& Main use cases& Characteristics& Operator\\  
\hline 
&\\

Bacs & Retail & Recurrent \& bulk payments; B2B payments, salaries, fees, utility bills, state benefits; often, long-term relationship between payer \& payee & £20 million limit,$^a$ high security, 3-5 days until clearing \& settlement, low fees (£0.05-0.5$^b$) & Pay.UK 
\\ & 
\\
FPS & Retail & One-off low-value payments; often consumers as payee or payer & £1 million limit,$^a$ immediate clearing, moderate fees (£1-5$^b$) & Pay.UK\\ & \\
CHAPS & Wholesale & Interbank market; high-value one-off purchases \& investments & No transaction limit, high security, immediate settlement, high transaction fees (£12-35$^b$) & Bank of England \\ & 
\\
ICS & Retail & One-off medium-high value payments; bills, warrants, travel cheques, payable orders & Cheque payments; mostly businesses & Pay.UK\\ & \\
LINK & Retail & Cash withdrawals and operation of ATM infrastructure & Consumer and business users; creates the link between physical cash and electronic book money & \\ & \\
Cards & Retail & One-off payments, dominant in consumer shopping (can be linked to mobile phones), international transactions possible & Systems completely run by single entity; moderate fees (2-6\% of transaction value plus additional charges)
& private \\
\hline 
&\\
SWIFT & Retail & International transactions in any currency by businesses and consumers & No legal limit,$^{a,c}$ up to 7 days for clearing \& settlement; moderate fees vary across banks and transaction types (£20-40$^b$ or 3-5\% of transaction value) & SWIFT\\  & \\     
SEPA & Retail & Transactions in EUR by businesses and consumers from and to the EU & Moderate fees (£1-20$^b$) & ECB \\  & \\  
TARGET2 & Wholesale & EU analogue to CHAPS; high-value transactions in EUR, mainly by businesses, used for transactions from and to the EU & High fees (£10-35$^b$) & ECB \\  & \\ 
\hline \\[-1.8ex] 
    \end{tabular}

    \vspace{0.25cm}
    
    \justifying \noindent
    Notes: The table reflects a time snapshot in 2023. A discussion of ongoing transformations and their expected impact is provided in \ref{app:subsec:innovation_payments_explained}. 
    $^a$ Banks may impose lower limits. $^b$ The fees are indicative and reflect approximate average values in 2023. 
    They vary across banks, transaction volumes, values, customers, and time, and banks may charge additional costs. Often, PSPs offer schemes that combine fix prices with percentage fees and price caps. 
    $^c$ High-value transactions are also regulated by anti-money laundering and tax policies. 
    
\end{table}

Payments in the UK can be made through different systems and payment instruments. Consumers and businesses use various systems depending on the type and purpose of a transaction. 
UK Finance reports 40.4 billion payment counts in the UK in 2021, whereby the majority are consumer payments. Businesses made about 5.5 billion payments, with 3 billion being B2B \citep{UKfinance2022UKpayments}. 
Our data covers only a small fraction (2.6\%) of all B2B transactions if measured as counts of executed transactions, but a significant amount of the transferred value (£$>$1.2 trillion in 2021). As a benchmark, the UK annual nominal GDP was £2.28 trillion in 2021.\footnote{Note that a direct comparison to GDP is not possible as the two variables are conceptually different.} 

Businesses often use different systems from consumers, depending on the transaction purpose, value, frequency, security, and costs. 
Table \ref{tab:overview_UKschemes} summarises the major payment schemes for electronic transactions in the UK, their main use cases, characteristics, and operators. 

The payment data used in this paper are a subset of Bacs transactions, one of the three major domestic schemes for B2B transactions in £, next to CHAPS and Faster Payment System (FPS). 
Bacs can only be used by businesses to initiate direct debit (DB) collections and direct credit (DC) transfers. To access Bacs services, businesses need to fulfil certain eligibility criteria. 

From a technical perspective, businesses can access Bacs services in three ways: (1) they can register their own Service User Number (SUN) and submit and receive payments themselves; (2) they can indirectly access the system via a so-called Bacs bureau while receiving their own SUN, but the bureau handles the transactions under their SUN; or (3) they rely on a third-party PSP that makes transactions on behalf of its customers using its own single SUN \citep{bacs2023guidedebit, bacs2023guidecredit}.

Compared to other payment options, Bacs transaction fees are very low (£0-0.5) and offer high security standards.\footnote{Transaction fees depend on the agreement between the PSP and the business. Fees usually vary across different account types and PSPs offering these services. In addition to fees, businesses also have to pay the set-up costs for obtaining a Bacs account.} 
Further, Bacs offers a relatively high transaction limit (£20m for businesses). 
This makes the scheme attractive for frequent and/or regular bulk payments. 
Bacs offers two payment instruments: Direct Debits (DD) and Direct Credits (DC), which are used for different purposes. Businesses use DC mainly for B2B payments and employment-related payments, such as payroll and pensions. The main use cases of DD in the B2B context are regular B2B collections, commercial billing, leasing, rental, and fee payments. 
DD provide a high guarantee to be paid on time \citep{payuk2013bacsprinciples}.  
Bacs DC and DD are also the major means of payment for governments to pay state benefits and collect taxes and national insurance contributions.
In \ref{app:sec:macro_benchmarking}, we provide some statistics and a short discussion about the differences in the economic information embodied in aggregate transactions of different payment instruments.

\begin{figure}

    \caption{Monthly time series of our payment data and major UK schemes}
    \label{fig:ts_UKpayments_benchmarking} 
    \centering
    \begin{subfigure}[b]{\textwidth}
        \includegraphics[width=\textwidth]{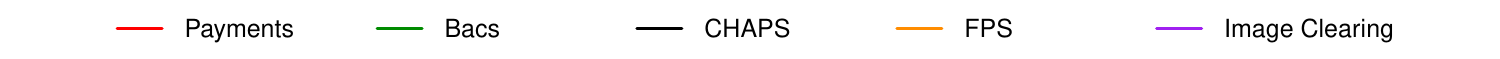}
    \end{subfigure}
    
    \begin{subfigure}[t]{0.49\textwidth}
        \centering     
        \includegraphics[width=0.8\textwidth]{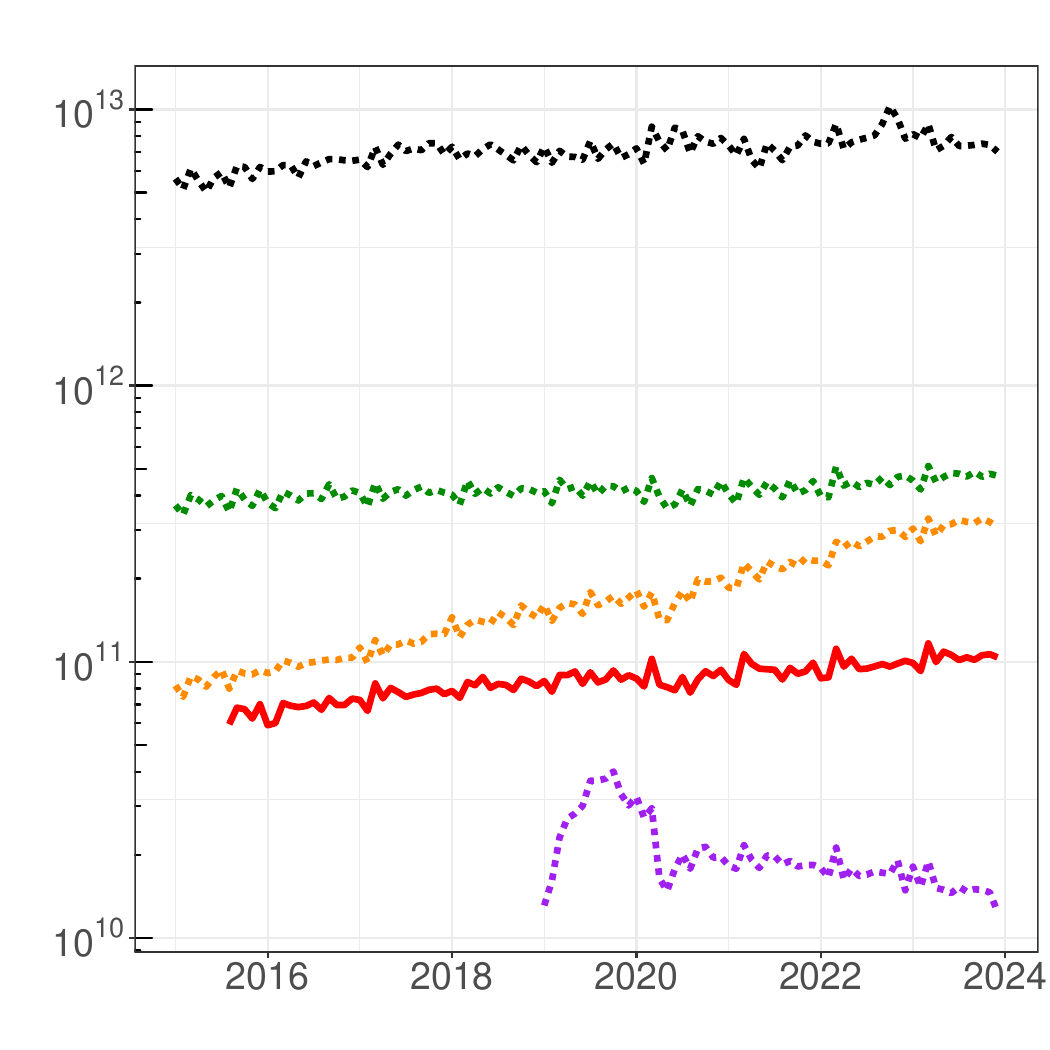}
        \caption{Values}
        \label{fig:ts_UKschemes_values}
    \end{subfigure}  
    \begin{subfigure}[t]{0.49\textwidth}
        \centering        
        \includegraphics[width=0.8\textwidth]{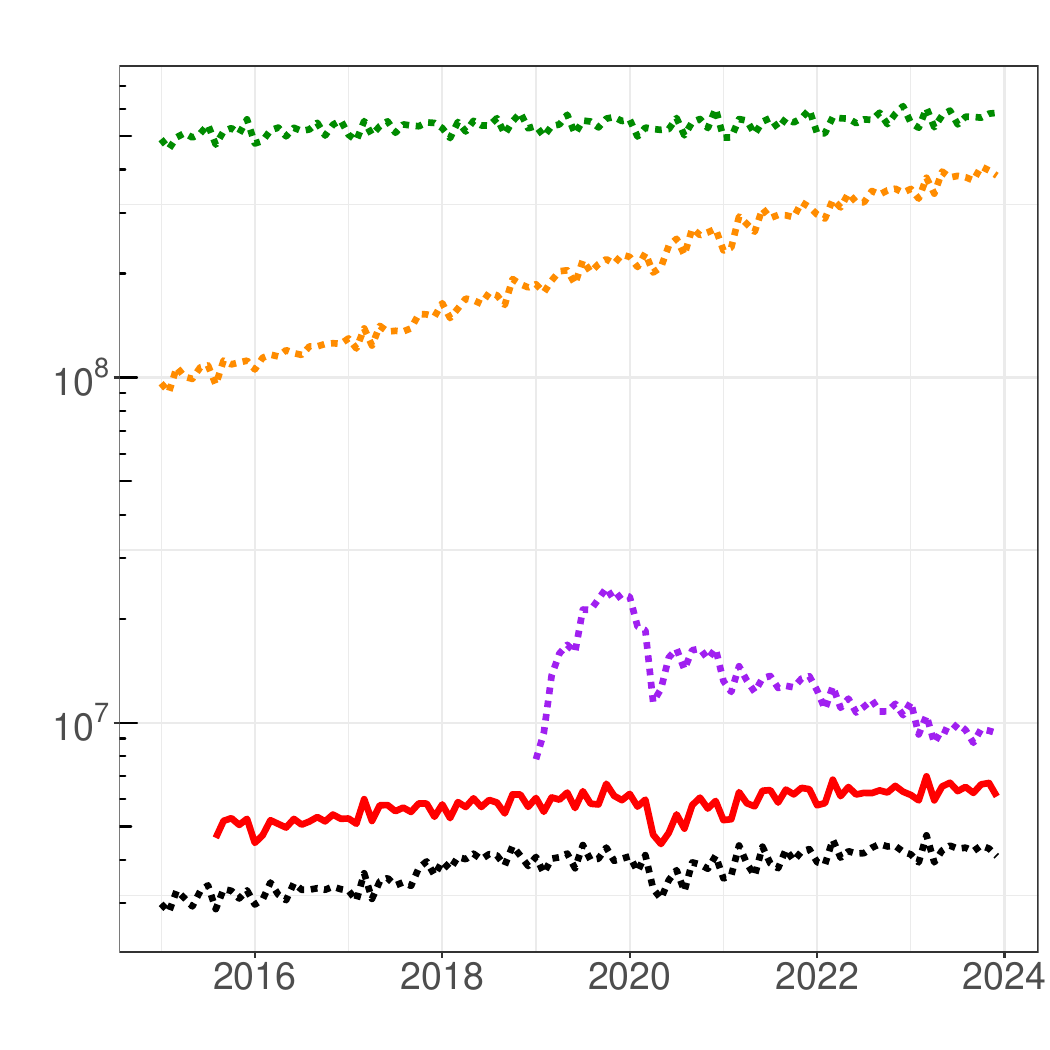}
        \caption{Counts}
        \label{fig:ts_UKschemes_counts}
    \end{subfigure}

        \justifying \scriptsize \noindent
    Notes: The vertical axis is scaled at a log-10 scale. Payments (red) are monthly aggregates of our data. The Bacs, CHAPS, FPS, and Image Clearing System data are downloaded from \citet{payuk2023historicaldata}. 
\end{figure}

The Bacs payment data used in this article are derived from an unweighted sample of anonymised and aggregated DD and DC payments between approximately 117,000 Bacs service users and capture roughly 22.1\% of the value of Bacs payments in 2023. The data set presents both the industry source and destination of the payments, with industries being assigned to SUNs using a combination of deterministic and probabilistic approaches matching Bacs service users' names to Companies House and other information \citep{ons2023interindustrymethods}.\footnote{A recent update to the data uses a significantly improved account classification approach that identifies more than 3 million organizations, with the greatest impact expected to be the inclusion of small and medium-sized enterprises in the data. Further improvements to the data are underway and are expected to become publicly available in the coming years. The new version of the data was released in the final stages of the publication process of this article and was therefore not included in the empirical analyses. Preliminary investigations indicate that the results are consistent, and the conceptual discussions and technical explanations provided in this article are expected to remain valid for future releases of these and similar data.}

Currently, the major alternatives to Bacs for electronic payments are the Faster Payment System (FPS), Card payments, and the Clearing House Automated Payment System (CHAPS). 
FPS, introduced in 2008, is the youngest of them and was a major payment innovation globally. It offers near-to-real-time clearing, which provides a high guarantee of payment. 
Compared to Bacs, the maximum transaction value for FPS is lower (£1 million) and the transaction fee is higher (£1-5).\footnote{The maximum transaction value was lifted from £250 thousand to £1 million in early 2023, and it is not yet possible to evaluate the impact of this increase on payment behaviour.} 
While still accounting for only a small share of annual payments by counts and values, the use of FPS has been increasing steadily (see Fig. \ref{fig:ts_UKpayments_benchmarking}), having reached an aggregate transaction value of almost £2 billion in the 2020s. 

Card payments are mostly used in consumer-to-business (C2B) transactions, especially in physical and online retail shopping. 
Transaction fees for card payments tend to be relatively high for businesses, while the exact conditions depend on the account type that businesses have with their PSP. 
CHAPS transactions are expensive for businesses and tend to be used only in special cases for high-value transactions, requiring a high security and eventually exceeding the transaction limit in the other schemes. 

The other domestic schemes are the Image Clearing System (ICS) for cheques and LINK, which connects electronic money to cash through withdrawals and cash deposits. Both are of minor and decreasing relevance, as suggested by the decreasing trends of cash and cheque usage for payments. 
UK businesses also use the international (SWIFT) and European schemes (SEPA, TARGET2), which are used mostly for international transactions in other currencies and therefore, the key payment channels for international trade. 

Fig. \ref{fig:ts_UKpayments_benchmarking} shows a time series of monthly aggregate transaction values and counts of our payment data and the other UK schemes (excluding cards), covering August 2015 to December 2023 using a log-10 scale. 

In 2023, the aggregate value of our payment data was £1.25 trillion, which corresponds to 22.1\% of the aggregate Bacs transaction values and 13\% when taking FPS, Bacs, and ICS together.\footnote{These numbers are calculated using the data after statistical disclosure control (SDC).}
The share of transaction counts is considerably lower (1.13\% for Bacs and 0.67\% for the aggregate). This can be explained by excluding consumer-related transactions, which are frequent but have a relatively low value compared to the transactions in our data. The average transaction value in our data was about £16.200 in 2023, which is about 20 times higher than the average Bacs transaction (£830). 

The values transferred through the CHAPS system are much higher. This is expected as it is a wholesale system for high-value transactions. CHAPS only indirectly reflects dynamics in the goods market but can be informative about the financial and interbank market, and potentially about investments. 

Over time (Fig. \ref{fig:ts_UKpayments_benchmarking}), the evolution of the aggregate transaction values in the payment data, Bacs, and CHAPS have been fairly stable, with minor monthly fluctuations and a moderate rise. FPS is the only scheme that exhibits a relatively steep rise over time in values and counts. ICS shows some fluctuations at the end of 2019, but a slowly decreasing trend reflecting the decreasing use of cheques. 

In summary, Bacs probably captures much of domestic intermediate trade in goods and services, yet coverage may vary across industries and types of transaction. In the long run, it may be influenced by the rise of FPS and other trends in payment systems (see \ref{app:subsec:innovation_payments_explained}). 

\FloatBarrier
\section{Macroeconomic benchmarking}
\label{sec:macro_benchmarking}
As a first step, we assess the economic information embodied in our payment data at the macroeconomic level by comparing it to GDP, monetary aggregates and other available payment data \citep{payuk2023historicaldata}.\footnote{We use chain volume GVA data as a proxy of GDP \citep{ons2023indicativeGDPdata}.}
Our results show that trends in the payment data behave similarly to those of Bacs totals and exhibit strong correlations with macroeconomic fundamentals, including GDP and monetary aggregates. 

\begin{table}[!h] \centering 
  \caption{Correlations with other payments and macro aggregates} 
  \label{tab:macro_benchmarking_3_noCovid} 
\scriptsize 
\begin{tabular}{@{\extracolsep{5pt}} ccccccccc} 
\\[-1.8ex]\hline 
\hline \\[-1.8ex] 
 & Bacs & FPS & CHAPS & GVA nsa & GVA sa & M1 nsa & M3 nsa & Prices \\ 
\hline \\[-1.8ex] 
 
\hline \\[-1.8ex] 
\multicolumn{8}{l}{\emph{Raw data in levels}}\\
\hline \\[-1.8ex] 
Yearly (value) & $0.967$ & $0.962$ & $0.926$ & $0.998$ & $0.998$ & $0.996$ & $0.948$ & $0.999$ \\ 
Monthly (value) & $0.874$ & $0.926$ & $0.794$ & $0.865$ & $0.915$ & $0.911$ & $0.921$ & $0.898$ \\ 
Yearly (count) & $0.972$ & $0.884$ & $0.949$ & $0.988$ & $0.996$ & $0.993$ & $0.969$ & $0.990$ \\ 
Monthly (count) & $0.817$ & $0.800$ & $0.934$ & $0.867$ & $0.854$ & $0.783$ & $0.806$ & $0.825$ \\ 
Yearly (avg) & $0.948$ & $-0.190$ & $-0.487$ & $0.992$ & $0.978$ & $0.979$ & $0.894$ & $0.991$ \\ 
Monthly (avg) & $0.696$ & $-0.095$ & $-0.409$ & $0.632$ & $0.858$ & $0.923$ & $0.919$ & $0.786$ \\ 

\hline \\[-1.8ex] 
\multicolumn{8}{l}{\emph{Growth rates}}\\
\hline \\[-1.8ex] 
Yearly (value) & $0.050$ & $-0.223$ & $0.462$ & $0.682$ & $0.066$ & $0.955$ & $0.787$ & $-0.837$ \\ 
Monthly (value) & $0.882$ & $0.695$ & $0.565$ & $0.720$ & $0.365$ & $0.579$ & $0.630$ & $0.189$ \\ 
Yearly (count) & $0.047$ & $-0.321$ & $-0.251$ & $0.686$ & $0.071$ & $0.956$ & $0.789$ & $-0.835$ \\ 
Monthly (count) & $0.619$ & $0.137$ & $0.382$ & $0.785$ & $0.328$ & $0.067$ & $0.138$ & $0.240$ \\ 
Yearly (avg) & $0.957$ & $0.732$ & $0.340$ & $0.415$ & $-0.261$ & $0.856$ & $0.616$ & $-0.905$ \\ 
Monthly (avg) & $0.667$ & $0.548$ & $0.138$ & $-0.162$ & $0.132$ & $0.738$ & $0.735$ & $-0.108$ \\ 
\hline \\[-1.8ex] 

\end{tabular} 

\justifying \noindent \scriptsize
Notes: 
This table shows Pearson correlations between annual (monthly) payments and other UK payment schemes and macroeconomic aggregates (GDP, M1, M3, Prices) during 2016 and 2023 (08/2015 and 12/2023), excluding the Covid-19 period, proxied by 2020 to 2022 (03/2020 to 12/2022). ``sa'' (``nsa'') is short for (non-)seasonally adjusted. Our payment data and other payment aggregates are compared by aggregate values, counts, and average values (short ``avg'') given by value divided by count. 
Growth rates are calculated as percentage growth compared to the (same month of the) previous year (for monthly data).\footnote{Bacs, FPS, and CHAPS data are obtained from \citet{payuk2023historicaldata}. 
Monthly GDP is proxied by indicative (non-)seasonally adjusted monthly ``Total Gross Value Added'' index data published by the ONS \citep{ons2023indicativeGDPdata, ons2023indicativeGDPadjusted}. 
``Prices'' is short for Consumer prices index data obtained from the OECD Key Economic Indicators (KEI) dataset \citep{oecd2023KEIdata}. M1 (M3) are narrow (broad) monetary aggregates, and thus nominal indicators, obtained from the OECD Main Economic Indicators (MEI) dataset \citep{oecd2023MEIdata}.}

\end{table} 

Table \ref{tab:macro_benchmarking_3_noCovid} shows monthly and annual correlations of our payment data with the other UK payment schemes, real GDP, monetary aggregates (M1) and prices, measured in levels (top rows) and growth rates (bottom rows). Data from the years of the Covid-19 pandemic (proxied by March 2020 to December 2022) are excluded. Additional results including Covid-19 are provided in \ref{app:sec:macro}. 
In levels, aggregate payment values and counts show strong correlations with the other UK payment data, ranging between 80-97\%. For transaction values, we find the highest levels for annual Bacs and monthly FPS aggregates, while CHAPS is highly similar in terms of counts. 

\begin{figure}[!h]
    \centering
    \begin{subfigure}[b]{0.9\textwidth}
        \includegraphics[width=\textwidth]{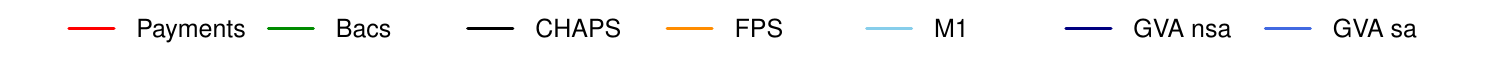}
    \end{subfigure}
    
    \begin{subfigure}[b]{0.32\textwidth}
        \centering
        \includegraphics[width=\textwidth]{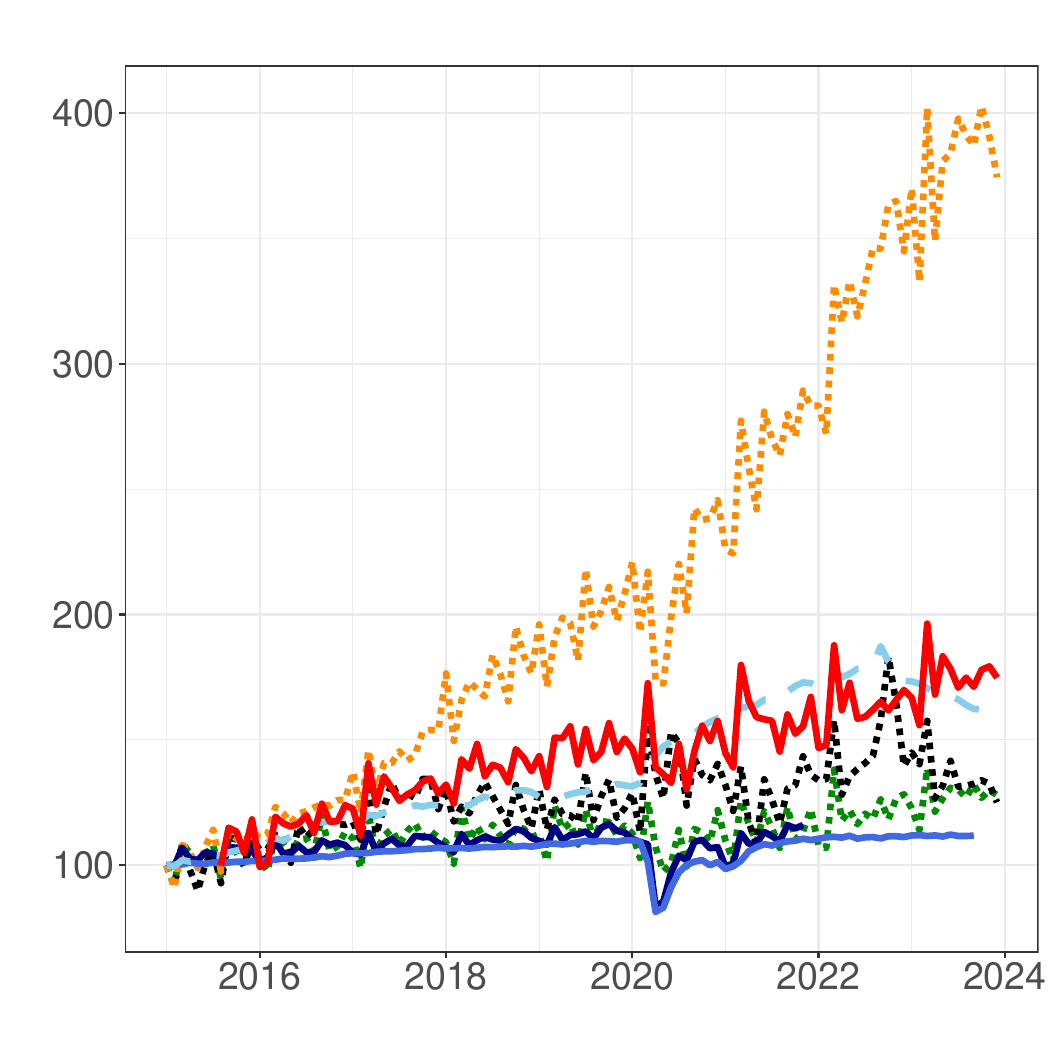}
        \caption{Value}
    \end{subfigure}
    \begin{subfigure}[b]{0.32\textwidth}
        \centering
        \includegraphics[width=\textwidth]{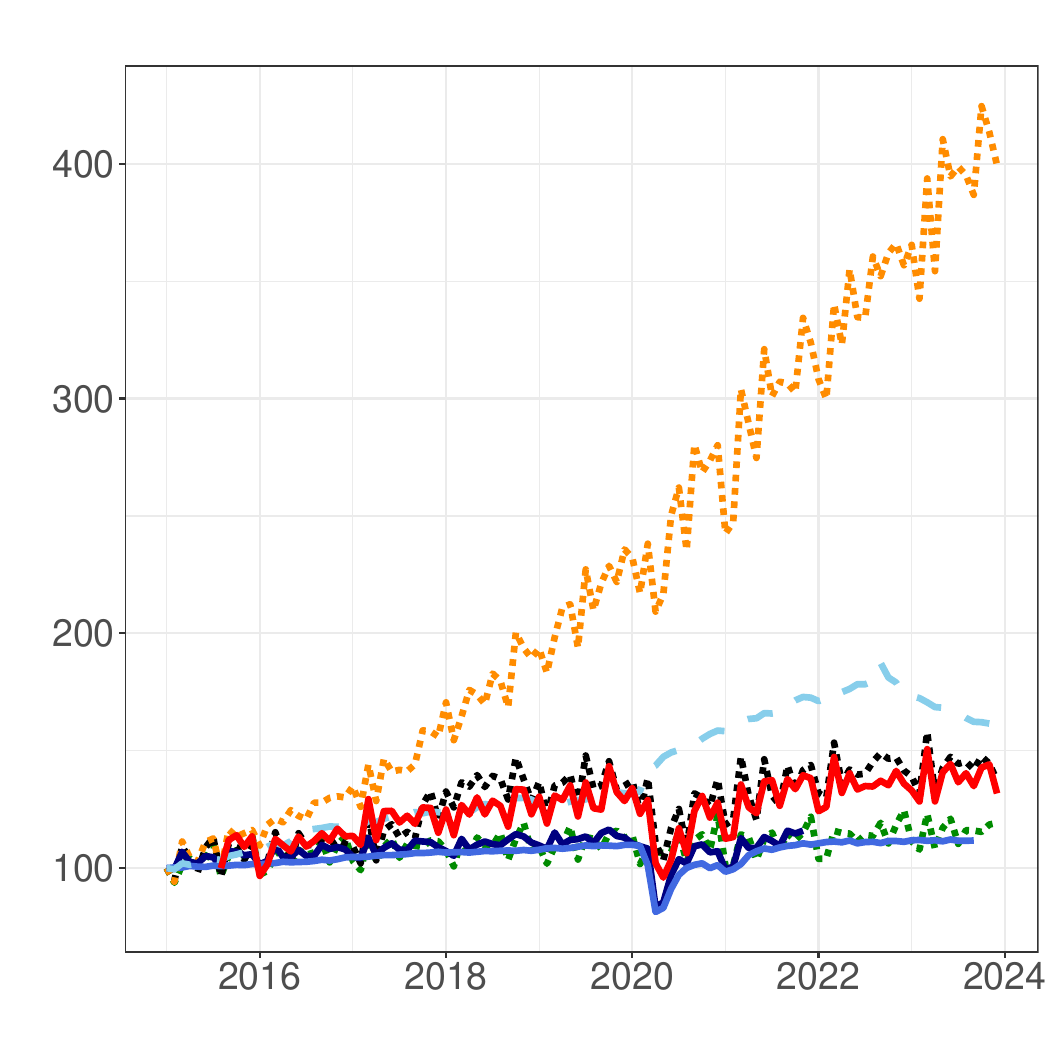}
        \caption{Counts}
    \end{subfigure}
    \begin{subfigure}[b]{0.32\textwidth}
        \centering
        \includegraphics[width=\textwidth]{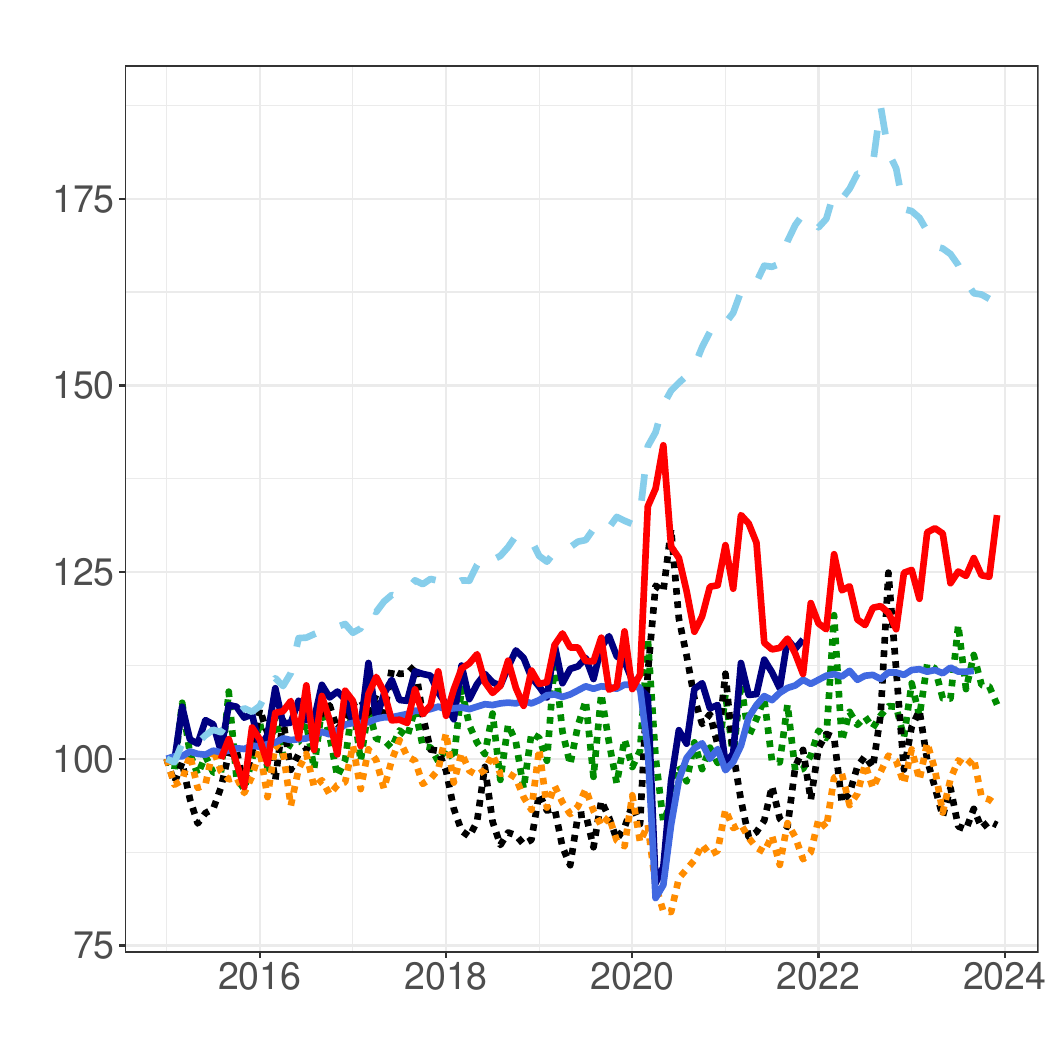}
        \caption{Average value}
    \end{subfigure}
    \caption{Monthly UK payments, GDP and M1}
    \label{fig:timeseries_aggr_index_payments_vs_gdp}

    \justifying \scriptsize
    \noindent
    Notes: These figures show monthly time series (indexed to 08/2015 = 100) for payments, the major UK payment schemes, and indicative (non-)seasonally adjusted monthly 'Total Gross Value Added' (GVA) data published by the ONS \citep{ons2023indicativeGDPadjusted, ons2023indicativeGDPdata}. Average values are obtained by dividing total values by counts.

\end{figure}

The growth rates exhibit more heterogeneous patterns: annual aggregates poorly correlate, which may be due to differences in the long-term trends (see also Fig. \ref{fig:timeseries_aggr_index_payments_vs_gdp}). In contrast, monthly growth calculated as growth in relation to the same month of the preceding year, shows fairly high correlations, especially for the Bacs value data with 88.2\%. 

The average transaction values show a high similarity with Bacs, but do not correlate with or only negatively correlate with the other payment schemes. Negative correlations of average values may indicate that the payment data capture different types of payments than those reflected in payment aggregates. For example, low-value payments in everyday expenditures differ from high-value investments or purchases of consumer durables. 

Looking at growth rates, we find higher levels of similarity, indicating that there may be a common underlying pattern of how transaction values evolve. One possible direction of interpretation may be their relationship with prices, here measured as the consumer price index. However, while finding strong positive correlations with prices measured in levels, we find a negative correlation when comparing by growth rates. This may seem counter-intuitive, but differences in trends may arise from sluggish price adjustments, especially when comparing consumer prices with B2B data.\footnote{We additionally made a comparison to producer price indices for manufacturing, but observed similar patterns.} 

Turning to economic fundamentals, we find strong correlations between payments and real GDP, ranging between 85-92\% for monthly data in levels. We analysed both seasonally adjusted (``sa'') and non-adjusted (``nsa'') data. 
The correlation performance for both indicators is similar.  
Looking at growth rates, the difference is more clear: at the monthly level, correlations between values (counts) are about 72\% (79\%) for non-adjusted data, but only half as high (36\% (33\%)) for seasonally adjusted GDP. These are very promising signals regarding the value of the data for applied economic research and advancing national statistics. 

As a next step, we relate payments to monetary aggregates, measured as M1 and M3, which can be considered an indicator of financial liquidity in the real economy.\footnote{M1 and M3 are monetary aggregates used as measures of the quantity of money and assets, while M3 includes assets at low levels of liquidity \citep{oecd2023MEIdata}.} 
Again, we observe strong statistical relationships for both M1 and M3 with high correlations of $>$90\% for payment values and around 60\% for their monthly growth rates. 
Correlations for payment counts are lower, with around 80\% for the data in levels and 7-14\% for growth rates. 

These observations confirm the idea of payment values as a nominal indicator and counts being more strongly related to data in real terms. 
B2B count data can indicate business dynamism: variations in the counts can indicate deviations from standing regular payments (such as fees, royalties, and loan repayments).

Fig. \ref{fig:timeseries_aggr_index_payments_vs_gdp} illustrates some of these observations, showing indexed monthly time series plots of real GDP, M1, payments, and other UK payment schemes for transaction values, counts, and the average value of transactions. 
We highlight five key observations: 
(1) By value, payments rose relatively more than GDP, CHAPS, and Bacs, and almost perfectly match with the long-term rise in nominal monetary aggregates M1 until 2022, when central banks began to tighten the money supply. 
By counts, the rise and fluctuations of the payment data almost perfectly co-evolve with CHAPS counts, and show very similar fluctuations as non-deseasonalised real GDP, but not the same long-term trend. 
(2) The Covid-19 shock in early 2020 shows ambiguous correlations: it is associated with a drop in GPD and payment counts, but peaking average transaction values and an unclear relationship to payment values (see also \ref{app:sec:macro}). 
(3) Average transaction values show the same pattern of growth as real GDP until the Covid-19 shock in 2020 when both time series decouple. The series re-converge over the following months showing a similar long-term trend. 
(4) The index series underline the steep rise of FPS. 
(5) Lastly, the time series shows some volatility, but no clear pattern of seasonality, yet higher correlations with non-deseasonalised GDP.

\FloatBarrier
\section{Comparison to national accounts}
\label{sec:benchmarking}
\FloatBarrier
Here, we first describe the construction of payment-based IOTs (Sec. \ref{subsec:harmonization_with_ONS}). Then, we compare the different IOTs by the network structure (Sec. \ref{subsec:aggregate_network}), auto- and cross-correlations (Sec. \ref{subsec:auto_cross_correlations}), and quantify edge-level differences (Sec. \ref{subsec:difference_quantification}). 

\subsection{From inter-industrial flow of funds to input-output tables}
\label{subsec:harmonization_with_ONS}

Payments in our data present both source and destination industries. They can be transformed into symmetric matrices of monetary flows, whereby rows are paying industries and columns are those being paid. Transposing the matrices leads to symmetric matrices of input-output flows, showing the row industries (being paid) as the input suppliers and column industries as (paying) customers. 

These matrices serve as proxies of IOTs, enabling a benchmarking exercise with the official NA tables. 
Here, we look at three different types of symmetric IOTs published by the ONS: (1) intermediate use within the supply and use tables (SUTs), and two analytical IOTs in an (2) industry-by-industry (IxI) and (3) product-by-product (PxP) format. These tables reflect supply chain linkages between industries.\footnote{For the SUTs, we used ``Table 2: Demand of products -- The 'Combined Use matrix' - Intermediate demand'', as published along with the ONS Blue Book \citep{ons2023blue}. This table represents the intermediate use of different products (rows) by industry (columns) as values of intermediate inputs, represented in purchasers' prices. The product classification is aligned with industry codes, which enables the symmetric representation as an IOT-proxy, while conceptual differences remain \citep{esa2010european}. Differently from the analytical IxI and PxP matrices, SUTs are available as a consistent time series, making them suitable for analyses over time.}

Industries are classified by SIC codes \citep{ons2009sic} and products by the Classification of Product by Activity (CPA) \citep{eurostat2015cpa}. These classifications are fully aligned with each other: at each level of aggregation, the CPA shows the principal products of the industries according to the SIC \citep[paragraph 9.2][]{esa2010european}.\footnote{The European standards refer to NACE (\emph{``nomenclature statistique des activités économiques dans la Communauté européenne''}) codes used at Eurostat, which are equivalent to the SIC used in the UK.} 

To compare the payments with the official IOTs, we aggregated monthly transactions into annual aggregates and harmonised the classification between the data sources. 
To construct the payment-based IOTs, we used data at the 3-digit level with 265 distinct industries for most sectors and 5-digit data with 612 different sectors whenever CPA codes were too granular for a 3-digit level matching. We applied this mixed procedure to maximise the coverage, as the statistical disclosure control (SDC) is more restrictive at the 5-digit level.\footnote{Appendix \ref{app:concordance} shows the mapping from SIC to CPA codes for each industry. The raw number of industry codes in the data is 705, but some of them were ``whitened'' due to the SDC.}  

We obtain a panel of annual proxy-IOTs covering the years 2016-2023.\footnote{The year 2015 is dropped due to incomplete coverage.} 
While our inter-industrial payment data include all Bacs payments received (limited by our data coverage), the ONS intermediate demand tables only cover payments received for an industry's primary product (see also Sec. \ref{sec:conceptual}). 
The official IOTs are compiled by the ONS in a stepwise procedure, whereby SUTs are the starting point. The SUTs show the flows of products and services in the economy across industries, products, and institutional sectors and with the rest of the world. The ONS assembles SUTs from a sample of almost 300 different data sources, consisting of business, and consumer surveys conducted annually by the ONS and other public and private datasets.\footnote{The list of data sources used for the SUTs is available here: \url{https://www.ons.gov.uk/economy/nationalaccounts/supplyandusetables/datasets/supplyandusetablesdatasourcescatalogue} [accessed on 2024/01/04]. See also \citet{ons2023blue}.} 
The data assembling follows international standards of balancing and applying national accounting identities \citep{esa2010european}. 

The intermediate demand within the SUT framework shows nationally supplied products and services plus imports used as production input for each industry, valued at current prices \citep[][ch. 9]{esa2010european}, excluding those contributing to gross fixed capital formation. 

The symmetric IxI and PxP tables are derived from the SUTs. They differ in how products and production activities are assigned to CPA codes. While intermediate demand within SUTs shows the use of products by industry, the symmetric tables show either how products are used to make products (PxP) or how the outputs of one industry are used as intermediate inputs in another industry (IxI) \citep[][par. 9.09]{esa2010european}. 
To simplify the language, we refer to the CPAs as industries, being aware that PxP tables and SUTs rely (partially) on products as units of analysis.

PxP tables focus on products that may be produced by various industries as their primary or secondary output, while IxI tables focus on industries that supply their primary output to multiple industries. Industries are classified by their primary production activity. 
The reallocation of non-primary products produced by an industry can be done in different ways, either by assuming that a certain product is always produced by using the same inputs, regardless of the industry producing it, or by assuming that a specific product is always sold to the same set of industries, regardless of the producer. UK IxI tables rely on the latter assumption. PxP tables are computed using both assumptions \citep{ons2023blue}.

The payment-based IOTs differ conceptually: they reflect transactions between multi-product industries, while the payment purpose remains unknown.\footnote{Theoretically, the trade flows in the payment data could be disambiguated, using a top-down imputation approach, where proportions are informed from other data sources. This was not tried for the existing data, also as there may be unknown issues, for example when primary and non-primary outputs are paid through other payment schemes. Working with the raw data on payment flows between multi-product industries can also be advantageous for certain applications: for example, diversification into new product markets can be an innovation strategy of firms and an indicator of technological and industrial change.} 
The industry classification is based on the self-declared business activity indicated as one or multiple 5-digit SIC codes when companies register at Companies House.\footnote{See \url{https://assets.publishing.service.gov.uk/government/uploads/system/uploads/attachment_data/file/1090526/IN01-V8.0.pdf} and \url{https://companieshouse.blog.gov.uk/2021/10/12/choosing-a-standard-industrial-classification-sic-code-for-your-company/} [accessed on 12/10/2023].}
In this experimental data version, we only rely on the first code indicated by the firms.\footnote{A recent update of the data relies on a classification that uses all reported SIC codes and assigns equal weight to payment flows. Relying only on the first code may introduce a bias toward over-representation of SIC codes with smaller numbers, as multiple codes tend to be listed in ascending order. Systematic checks for possible biases were beyond the scope of this article.}

Further, when comparing official NAs to the payment data, some product categories are entirely missing in the payments, such as ``T97 - Activities of households of domestic personnel'' or ``Imputed rents'', which is a natural feature of a dataset based on B2B payments. 

We compare the payment-based IOTs using both transaction counts and values with the SUT, IxI, and PxP tables. 
The data availability varies: payments are available for 2016-2023, IxI for 2018-2019, PxP for 2010, 2013-2015, 2017-2019, and SUT for 1998-2021.\footnote{We exclude the PxP from 2016 as they rely on another industry disaggregation and cover only 64 sectors. The SUT series is taken from the ONS Blue Book 2023 \citep{ons2023blue}. All data has been downloaded from the ONS website in Q1/2024.} 
The availability of the official data reflects the publication delays caused by the complex data collection and compilation procedure when merging and harmonising data from heterogeneous sources.\footnote{In the Blue Book 2023 \citep{ons2023blue}, the catalogue of data sources used for the SUTs includes 279 different entries, including data from public institutions like the ONS, the BoE, Treasury, Tax and Customs offices, government departments, private sector-specific data providers, international institutions, data from other public institutions and subnational authorities.} 
The compilation of the official tables is occasionally revised in response to economic change and methodological improvements. Only the SUTs are revised backwards, thus providing a consistent time series. However, inconsistencies can still arise from improvements and extensions of the data collection process, for example when surveys are amended.

\subsection{Aggregate network statistics}
\label{subsec:aggregate_network}
We now analyse the IOTs from a network perspective, representing the tables as weighted networks of industries trading goods and services. The network view is relevant as most supply chain and input-output analytics rely on network methods \citep{carvalho2014from, acemoglu2012network, leontief1991economy, roson2016input}. 
The nodes in the network are given by the industries (CPA codes) and the links are transactions between two industries. The links are weighted by the transaction value (or count) $Z^{\alpha}_{ij}$ between two industries $i$ and $j$, where $j$ buys inputs from $i$. As a notation, we use $\alpha$ to indicate the type of IOT with $\alpha = \{ \text{Value}, \text{Count}, \text{IxI}, \text{PxP}, \text{SUT}\}$.  

We also calculate input (output) shares $\omega^{\text{in}}_{\alpha,ij}$ ($\omega^{\text{out}}_{\alpha,ij}$) by dividing the raw weight of an input (output) link $Z^{\alpha}_{ij}$ ($Z^{\alpha}_{ji}$) by the sum of inputs purchased (outputs sold) from (to) all other industries, given by

\begin{align}
        \omega^{\text{in},\alpha}_{ij} = \frac{Z^{\alpha}_{ij}}{\sum_{i} Z^{\alpha}_{ij}} &&
        \left(\omega^{\text{out},\alpha}_{ij} = \frac{Z^{\alpha}_{ji}}{\sum_{j} Z^{\alpha}_{ij}}\right). 
        \label{eq:shares_in_out}
\end{align}

Using this network interpretation, we calculate aggregate properties of the payment-based and ONS IOTs, shown in Table \ref{tab:network_stats_2019_no_truncation} using 2019, which is the most recent year for which IOT data were available when writing the paper (Q1/2024). 
The upper part of the tables illustrates the statistics when using raw transactions as weights, the lower parts when using input and output shares. 

\begin{table}[!htp] \centering 
  \caption{Properties of the payment and ONS input-output networks in 2019} 
  \label{tab:network_stats_2019_no_truncation} 
\footnotesize 
\begin{tabular}{@{\extracolsep{8pt}} lccccc} 
\\[-1.8ex]\hline 
\hline \\[-1.8ex] 
 & Value & Count & PxP & SUT & IxI \\ 
\hline \\[-1.8ex]
\underline{\emph{Raw transactions} }\\ 
Density & $0.286$ & $0.286$ & $0.723$ & $0.474$ & $0.980$ \\ 
Average degree & $28.550$ & $28.550$ & $75.202$ & $49.260$ & $101.885$ \\ 
Average strength & $2,783.139$ & $239,906.400$ & $10,563.500$ & $12,741.830$ & $10,593.480$ \\ 
Average weight & $97.483$ & $8,403.027$ & $140.468$ & $258.667$ & $103.975$ \\ 
Reciprocity & $0.554$ & $0.554$ & $0.793$ & $0.534$ & $0.989$ \\ 
Transitivity & $0.648$ & $0.648$ & $0.921$ & $0.787$ & $1$ \\ 
Assortativity by degree & $-0.358$ & $-0.358$ & $-0.176$ & $-0.190$ & $-0.005$ \\ 
 \hline \\[-1.8ex] \underline{\emph{Input shares}} \\  
Average strength & $0.885$ & $0.940$ & $0.840$ & $0.741$ & $0.846$ \\ 
Average weight & $0.031$ & $0.033$ & $0.011$ & $0.015$ & $0.008$ \\ 

\hline \\[-1.8ex]  \underline{\emph{Output shares}} \\ 

Average strength & $0.839$ & $0.864$ & $0.812$ & $0.731$ & $0.828$ \\ 
Average weight & $0.029$ & $0.030$ & $0.011$ & $0.015$ & $0.008$ \\ 
\hline \\[-1.8ex] 
\end{tabular} 

\vspace{0.25cm}

\justifying \noindent \scriptsize
Notes: The first (second) column uses payment values (counts) as weights. The other columns represent official IOTs published by the ONS, where PxP is short for Product-by-Product, IxI for Industry-by-Industry, and SUT for Supply-and-Use Table. The data are aggregated into 105 distinct CPA codes (see Sec. \ref{subsec:harmonization_with_ONS}).
Raw transaction data are shown in £ million. 

\end{table} 

The comparison reveals that the payment networks are much less densely connected than the ONS SUT, PxP, and IxI networks, reflected in a low density of $<$30\%, compared to 47-98\% for the ONS tables. 
On average, an industry has about 29 payment links out of 105 theoretically possible links, as reflected by the degree, including loops reflecting within-industry trade. 
The connectivity is strongest in the IxI network which is almost fully connected, and lowest in the SUT, with a density of 47\% and 45 links per industry on average. 
\begin{figure}[h]
    \centering

    \caption{Network density at different truncation thresholds}
    \label{fig:network_density_plot_3digit}

    \begin{subfigure}[b]{\textwidth}
        \centering
        \includegraphics[width=\textwidth]{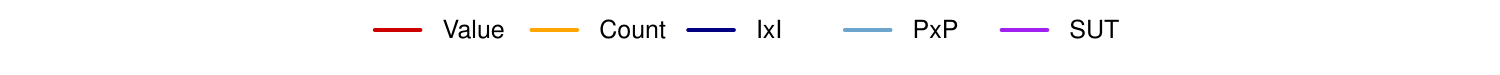}
    \end{subfigure}

    \begin{subfigure}[b]{0.44\textwidth}
        \centering
        \includegraphics[width=\textwidth]{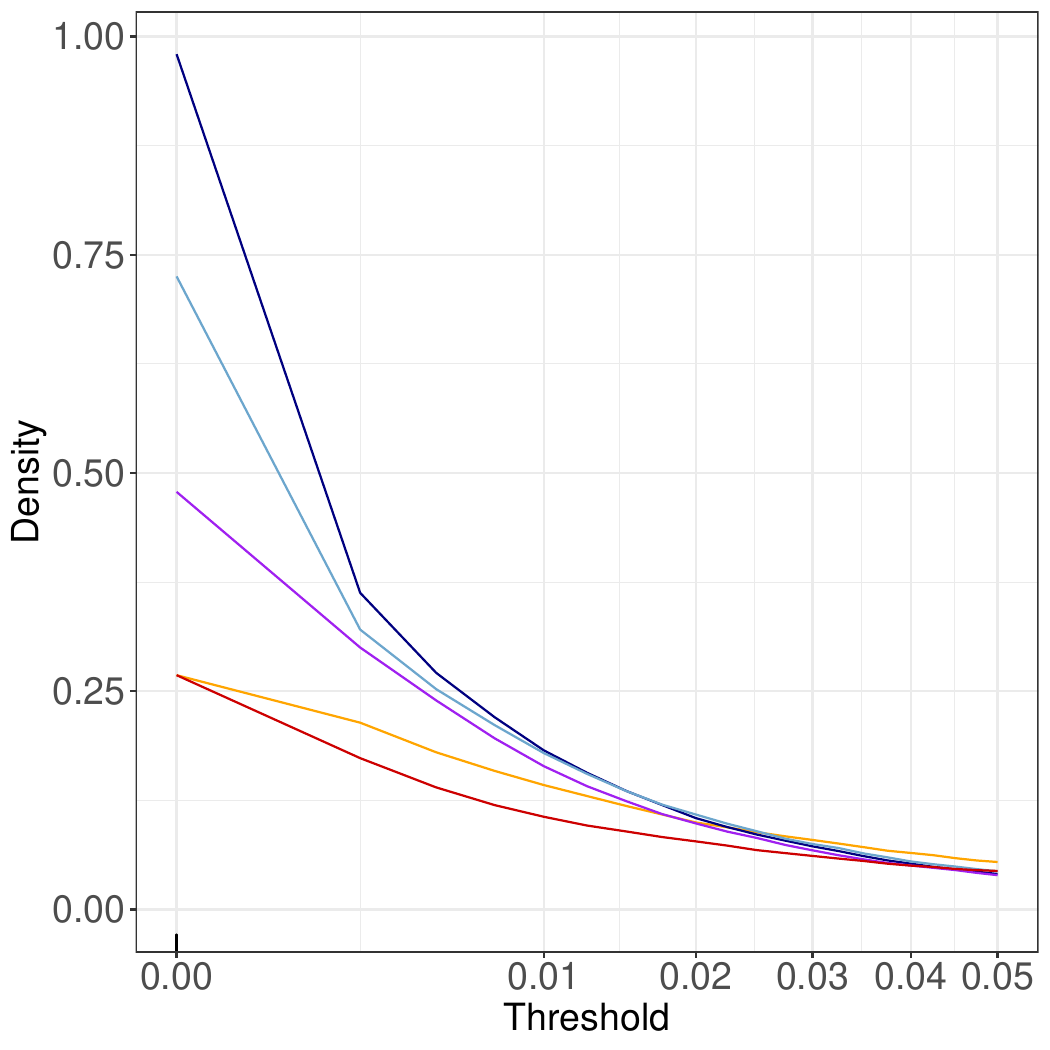}
        \caption{Input network}
    \end{subfigure}
    \begin{subfigure}[b]{0.44\textwidth}
        \centering
        \includegraphics[width=\textwidth]{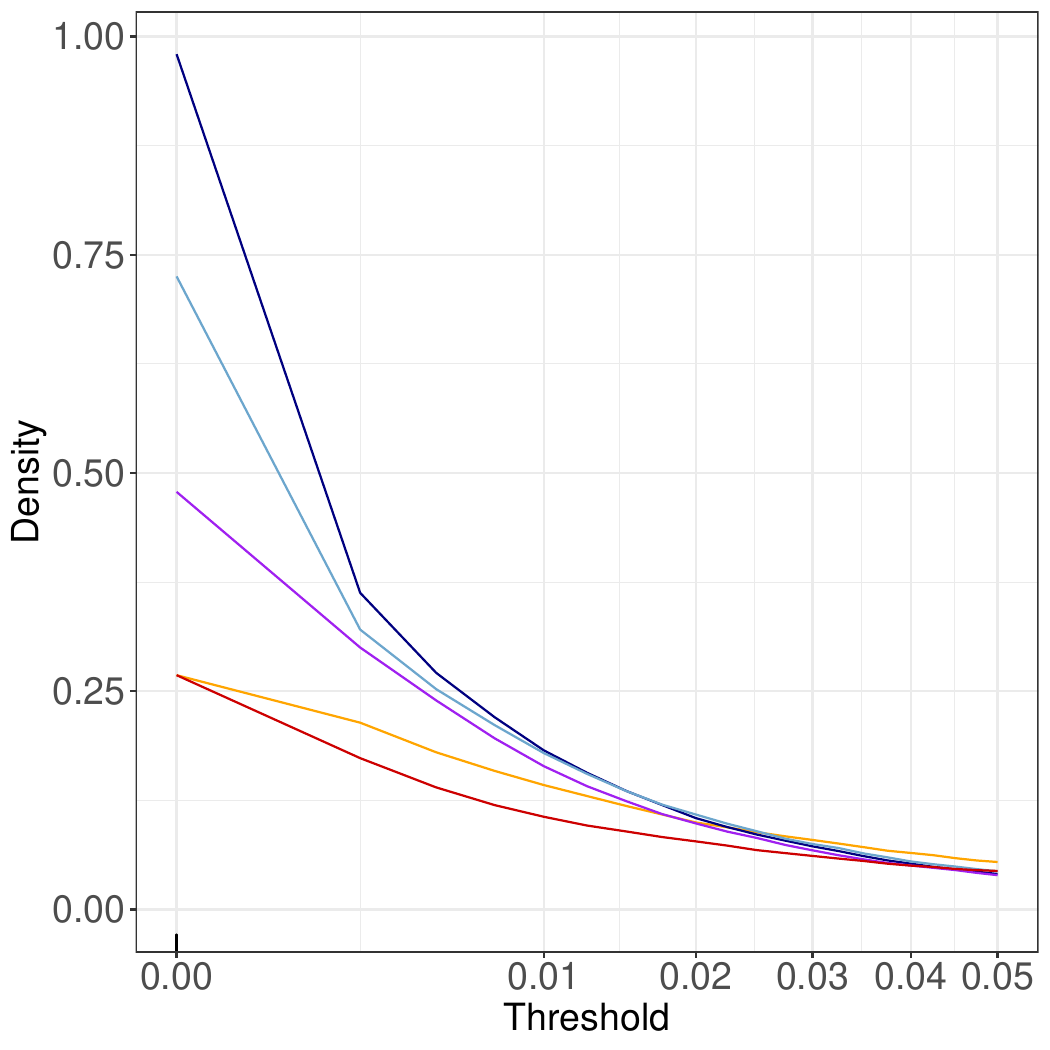}
        \caption{Output network}
    \end{subfigure}

    \justifying \scriptsize \noindent
    Notes: This figure shows the effect of network truncation thresholds (x-axis) on the network density (y-axix). In the left (right) figure, a link is removed if the input (output) share is smaller than the threshold value. 

\end{figure}

Fig. \ref{fig:network_density_plot_3digit} illustrates how the density decreases in the ONS-based IOTs if we truncate the network by removing non-significant links: for example, if we remove all links in an IOT $\alpha$ with an input (output) share $\omega^{\text{in},\alpha}_{ij}$ ($\omega^{\text{out},\alpha}_{ij}$) smaller than 1\% (5\%), the density drops from 47-98\% to 16-18\% (3-4\%). The payment-based IOTs are less sensitive to such truncation, declining from 29\% to 11-14\% (4-6\%) and become even denser than the ONS IOTs if the truncation is strong (see also \ref{app:subsec:aggregate_network}). 
This arises from the SDC procedure, where all small and potentially disclosive links between industries have already been removed.\footnote{This article relies on an experimental version of the data. A recent data update included an expanded coverage of firms and further improvements are underway. This will likely contribute to a higher connectivity of the non-truncated network.}

Transitivity and reciprocity range between 55\% and 100\%, and the higher the number the denser is the network. 
This number indicates the share of industries, which are customers and sellers to each other at the same time (reciprocity) or are connected through a third industry, forming a closed triad (transitivity). 
These values are lowest for the SUT and the payment-based IOTs.  

All networks show a negative node assortativity, telling us that large and well-connected industries tend to trade more with smaller and less connected industries, whereby ``well-connected'' means a high number of links (degree), and ``large'' refers to a high level of output or input. 
The negative assortativity of the raw networks is an often documented property of IOTs \citep{hotte2023demand}, but not surprising given the high density. 
The assortativity becomes positive if we impose a network truncation of 5\% (see Table \ref{app:tab:network_stats_2019_threshold_pct_5}), meaning that large industries are more frequently connected to other large sectors. This change in truncation thresholds is qualitatively consistent across all IOTs.  

The payment-based IOT proxies and official IOTs are qualitatively consistent by network properties. We found consistent responses in almost all indicators when removing links with a small economic weight, but the lowest sensitivity to truncation for the payment-based IOT, likely arising from SDC. At the 5\% truncation level (often imposed on IOTs before studying spillovers \citep[see][and references therein]{hotte2023demand}), the aggregate network properties of IOTs from the different data sources are very similar.  

Beyond the impact of the SDC, the quantitative variations and their sensitivity to link removal can be associated with differences in the compilation procedure and data sources.  
The payment data capture every financial transfer between two businesses, which can produce multiple products and services, leading to an aggregation into multi-product industries. 
In contrast, the ONS tables are based on surveys asking businesses to state their major purchases, mostly by product category. The assignment of product categories to industries is associated with several steps of harmonisation and balancing.
The SUTs are the ``rawest'' form, showing which industry used which products as intermediate inputs. The SUTs do not reflect whether the products are the inputs to or outputs of primary or non-primary production. 
This explains lower connectivity in the SUTs: including non-primary production implies an imputation of additional links, leading to a higher density of the PxP and IxI networks. 

Also, the treatment intermediary industries (trade, retail, finance) plays a role, as the payment data show the full transfer as a transaction from or to the intermediary and ONS tables only allocate the margin charged on the service provided to the intermediary, but add a new link between the seller and final user of the traded good, leading to a higher density and transitivity of the networks (see also Sec. \ref{sec:conceptual}).  

\FloatBarrier
\subsection{Auto- and cross-correlations}
\label{subsec:auto_cross_correlations}
\begin{figure}[!h]

    \caption{Auto- \& cross-correlations of input and output shares (2018-2019)}        
    \label{fig:correlations_edge_level_ONS_payments_IO_shares}
    
    \centering
    \begin{subfigure}[b]{0.49\textwidth}
        \includegraphics[width=\textwidth]{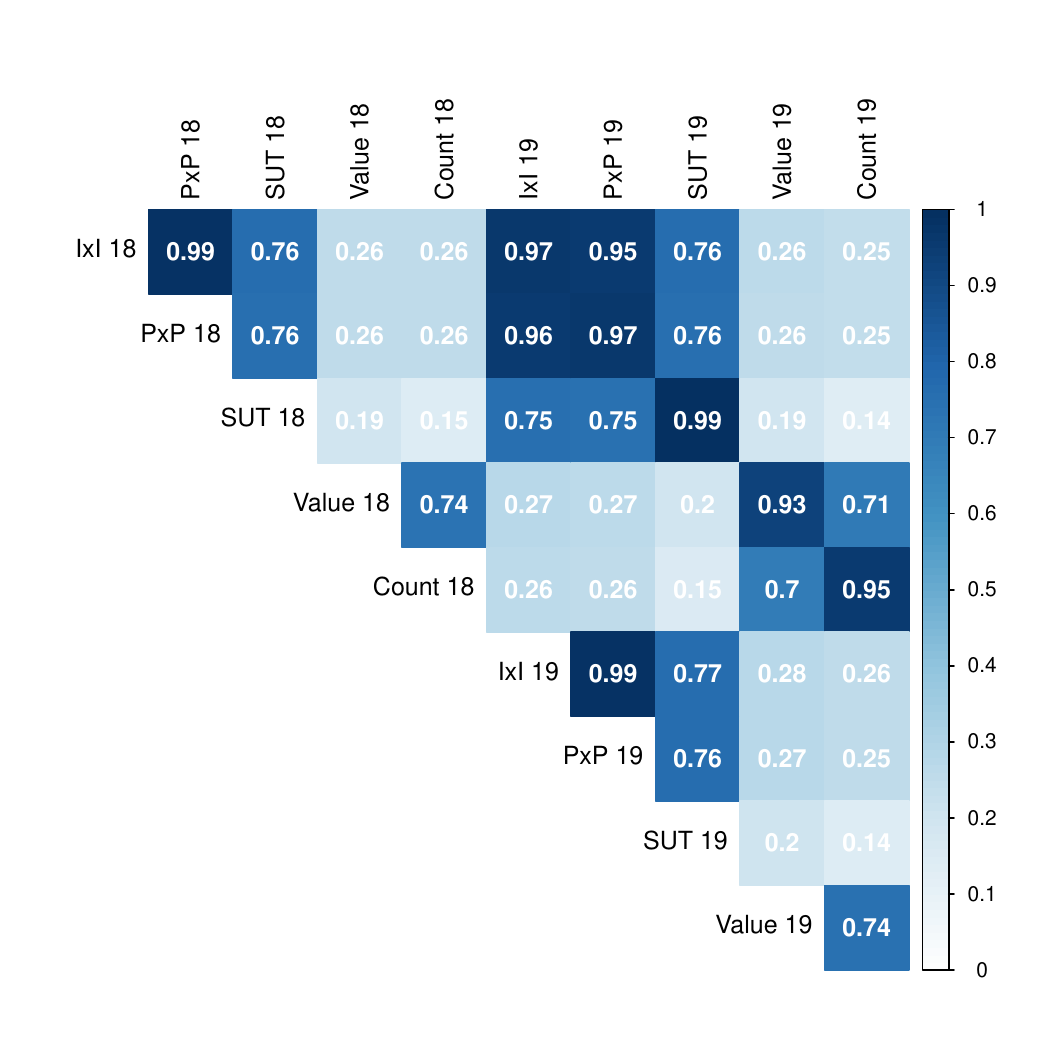}
        \caption{Input shares}
        \label{subfig:correlations_edge_level_ONS_payments_input_shares}
    \end{subfigure}
    \begin{subfigure}[b]{0.49\textwidth}
        \includegraphics[width=\textwidth]{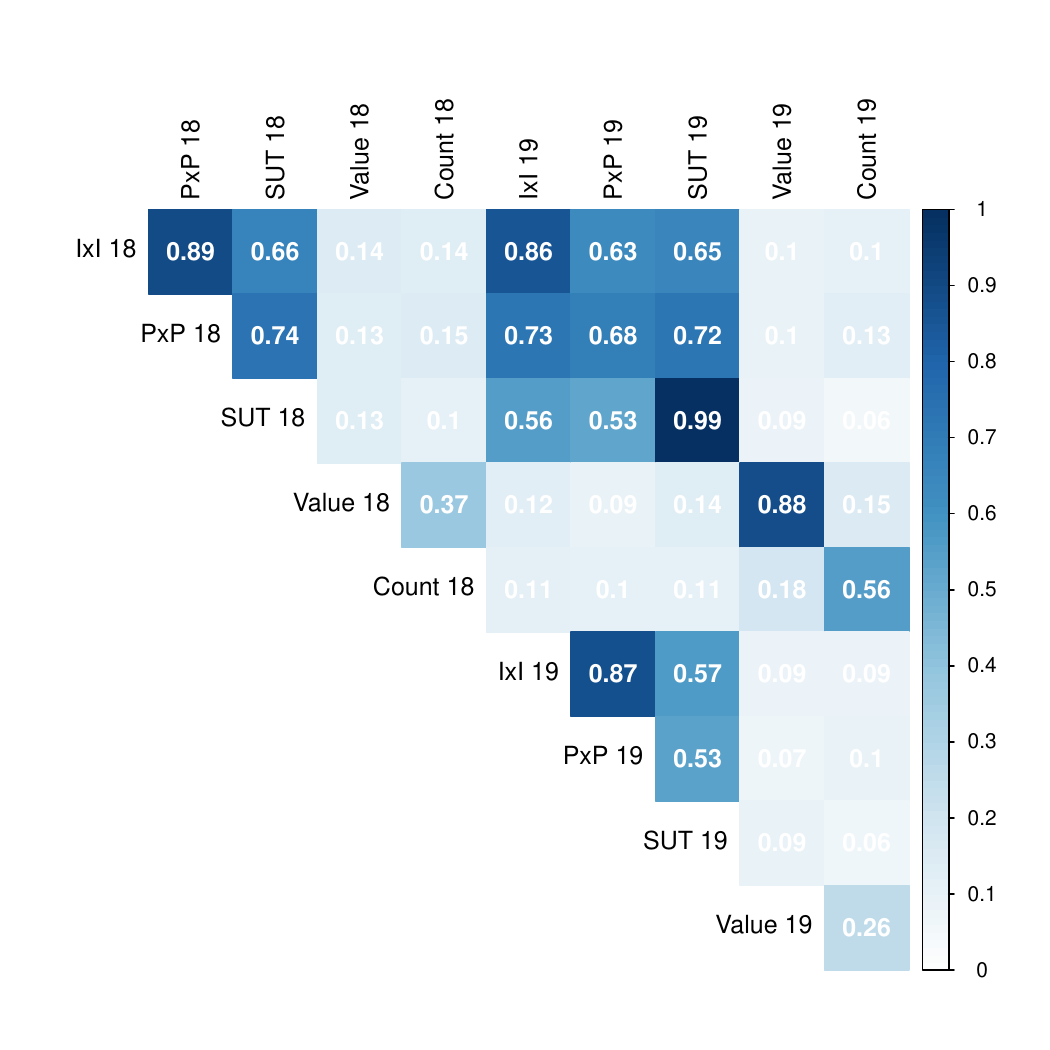}
        \caption{Output shares}
        \label{subfig:correlations_edge_level_ONS_payments_output_shares}
    \end{subfigure}

    \justifying \scriptsize \noindent
    Notes: The correlations are measured by the Pearson correlation coefficient between input and output shares in the payment-based IOTs (values and counts) and the IxI, PxP, and SUTs. 
    
\end{figure}

Next, we analyse auto- and cross-correlations of IOTs at the edge , and industry levels, exploring how inputs and outputs of industries auto-correlate within the same IOT and cross-correlate across different IOTs. The edge-level results are illustrated by Fig. \ref{fig:correlations_edge_level_ONS_payments_IO_shares}, which shows pairwise auto- and cross-correlations of in- and output shares in the different IOTs from 2018-2019 using the Pearson correlation coefficient. A darker colour indicates stronger correlations. 

The similarities of the IOTs across data sources (ONS, payments) by input shares are higher than similarities by output shares, ranging between 14-28\% instead of 6-15\%. 
Generally, we observe high within-data source cross-correlations, with coefficients for the input side ranging between 76-99\%, and high auto-correlations within the same IOT.  
Similarities are higher at the input rather than the output side. This decline is strongest for the within-payment correlations between values and counts (declining from 74\% to 36-37\%), and for the similarity of SUTs to the analytical IxI and PxP tables, going down from $>$75\% to 50-70\%. 
These findings are consistent with an analogous correlation analysis at the industry level, comparing industries by aggregate inputs and outputs, as illustrated and discussed in \ref{app:subsec:auto_cross_correlations}. 

The analysis reveals three core insights: (1) Within-ONS and within-payment similarities are larger than across data sources for any measure and across time. (2) We find very high auto-correlations (up to 99\%). (3) At the edge level, similarities by input links are much higher than outputs. At the industry level, aggregate inputs in the payment-based IOTs are most similar to SUTs, but more similar to PxP and IxI by output. 
This might be explained by the nature of SUTs, reflecting the supply of an industry classified by its primary output. In contrast, the input side of SUTs is based on the correct classification of products used as inputs of multi-product industries.

Relatively high similarities of the payment-based and ONS IOTs at the industry and transaction level, without any pre-processing or statistical data cleaning, are promising signals for using the data in applied economic research at the macro, industry, and network level.

\FloatBarrier
\subsection{Quantifying the edge-level difference}
\label{subsec:difference_quantification}
After analysing similarities, we now quantify typical differences. 
We face three issues: (1) the overall value of transactions is different across datasets, caused by the under- and over-sampling of industries (see also \ref{app:scale_differences} and \ref{app:tab:top10_influence_vector}). This undermines the direct comparison. To improve the comparability, we rescale the values. 
(2) Some industry pairs have much higher values of mutual transactions than other pairs. Therefore, the difference between the ONS and payment data tends to be extremely high for industry pairs with very high transaction values. To solve this issue, we measure of relative differences in absolute value.
(3) Because there are cases where one of the two datasets has a value of precisely zero. Hence, we only compare those transactions between industries, which are non-zero in both datasets.

We develop a measure, called proportional difference, indicating how many times larger a value is in one dataset compared to the other: a value of 1 means that the two values, measured as a proportion of the total transaction value, are equal (zero error), and a value of 2 means that the value is twice as large in one dataset compared to the other. Details are provided in \ref{app:subsec:diff_quantification}.

\begin{figure}[h]
    \centering
    \caption{Proportional differences between the ONS and payment-based IOTs}
    \label{fig:rel_diff_histogram_positives}

    \includegraphics[width=0.9\textwidth]{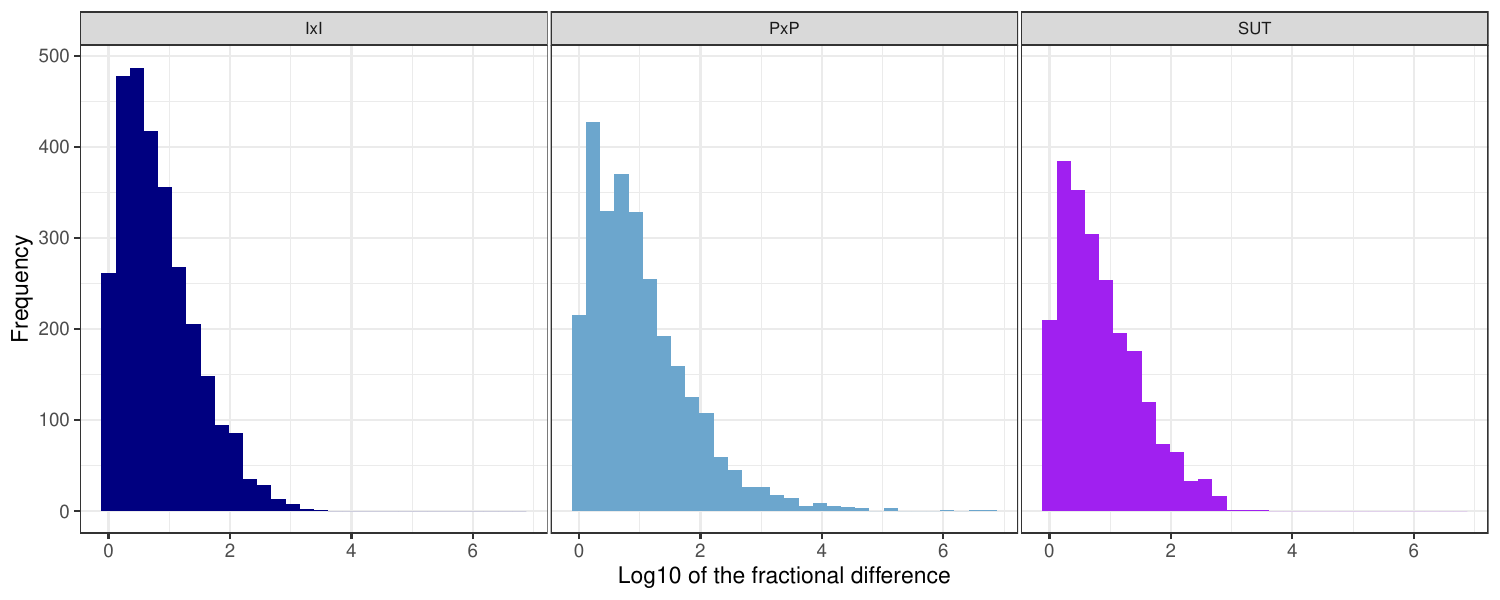}

    \justifying \scriptsize
    \noindent
    Notes: The figures show the distribution of the proportional edge-level differences (scaled at a log-10 basis) between the payment-based and ONS IOTs, using 2019 data. Industry pairs are removed if the transaction value is zero in one of the two datasets.\footnote{A comparison of differences with and without link removal are available in \ref{app:subsec:diff_quantification}, using a scaled percentage difference metric.} 
\end{figure}

\begin{table}[h]
\centering
\caption{Quantiles of the proportional differences}
\begin{tabular}{p{1.75cm}p{1.5cm}p{1.5cm}p{1.5cm}p{1.5cm}}
\hline \hline \\[-1.8ex] 
             & 25\% & 50\% & 75\% & 100\% \\ 
\hline \\[-1.8ex] 
IxI  & 2.20 & 5.10 & 15.24 & 3851.66 \\
PxP  & 2.40 & 6.76 & 27.74 & 582092.2  \\
SUT  & 2.12 & 5.08 & 17.01 & 2911.96 \\

\hline \hline \\[-1.8ex]
\end{tabular}

   \justifying \scriptsize
    \noindent
    Notes: Quantiles of the proportional differences between the IxI, PxP, SUTs and the payment-based IOT in 2019. Unlike as in Fig. \ref{fig:rel_diff_histogram_positives}, the values are not log-scaled. 
    \label{tab:quantiles_diff}
\end{table}

Fig. \ref{fig:rel_diff_histogram_positives} shows the histograms of proportional differences, scaled to a log-10 basis. The quartiles of the distributions are summarised in Table \ref{tab:quantiles_diff}. 

The results show that the differences can be extremely large: the medians range between 5.1-6.8 for all the three IOTs, meaning that for half of the pairs, one value is at least 5-7 times larger than in the payment-based IOT. 
The 25\% and 75\% quantiles range around 2.1-2.4 and 15.2-27.7, indicating a highly skewed distribution with long tails. For most industries, we observe moderate deviations, but for some industry pairs, the differences are extreme. 
Due to the different coverage of the two data sets and the proportion of data not allocated to any industry in the payment data (about 40\%), this is not a surprising result.\footnote{upcoming data releases with improved classification methods will probably show significant advances in coverage.}  

Further, and consistent with earlier results, the differences are largest between the payments and the PxP table, with the most extreme outliers. The differences between payments and the IxI and SUT range around similar values, while the distribution for IxI tends to have a thinner tail. 

The restriction on quantifying the differences between non-zero transaction links in both datasets may be seen as a distortion. In \ref{app:subsec:diff_quantification}, we show additional results for an alternative metric that allows keeping the one-sided zero entries. We observe that the differences tend to be slightly smaller and less skewed, but the effect is small.

\FloatBarrier

\section{Conceptual differences between payments and National Accounts}
\label{sec:conceptual}
Now, we discuss the main conceptual differences between payment-based and official IOTs, focusing on the (likely) most impactful aspects summarised in Table \ref{tab:conceptual_diffs}. This section is partly based on \citet[Appendix A]{bacilieri2022firm}, with a similar discussion of firm-level production networks constructed from VAT data. We supplement our discussions with information on Bacs processing statistics \citep{payuk2021statistics}. 

\begin{table}[!ht]
\centering

\caption{Conceptual differences}
\label{tab:conceptual_diffs}

\begin{tabular}{@{\extracolsep{8pt}}| p{0.175\textwidth} | p{0.355\textwidth} | p{0.355\textwidth} |}
\hline \hline &&\\[-1.8ex]
\textbf{National accounts element} & \textbf{SUT Intermediate Use} & \textbf{Payments} \\
\hline \hline &&\\[-1.8ex]
Time of recording & Products enter the production process & payment takes place \\
\hline &&\\[-1.8ex]
Gross fixed capital formation & Excluded & Likely included; debt repayment for financing GFCF also likely included \\
\hline &&\\[-1.8ex]
Financial services & Included & Flow of funds to financial sector likely to include payments from debtor to creditor, including financial services\\
\hline &&\\[-1.8ex]
Goods and services bought for resale & Excluded & Likely included \\
\hline &&\\[-1.8ex]
Distributive transactions & Excluded; taxes are added to basic prices while subsidies are deduced to calculate purchasers’ prices  & Likely partially included, examples can include dividends and interests, insurance premiums and settlements, taxes and subsidies\\
\hline &&\\[-1.8ex]
International trade & Exports excluded 
Imports included& Likely excluded \\
\hline &&\\[-1.8ex]
Inventories & Excluded  & Likely included \\
\hline &&\\[-1.8ex]
Industrial classification & Reported by firms answering ONS surveys and other data sources & Results of matching exercise from SUNs to Companies House data and other information\\
\hline \hline

\end{tabular}
\vspace{0.125cm}

\justifying \scriptsize
    \noindent
    Notes: This table summarises the conceptual differences, focusing on the intermediate consumption table obtained from the SUTs (``Demand of products -- The `Combined Use matrix''). It represents industries' intermediate demand at purchasers' prices and is published as part of the SUTs available in the ONS Blue Book \citep{ons2023blue}. Most of the issues are equally valid for the analytical IxI and PxP tables. 
\end{table}

\paragraph{Time of recording and inventories}
NAs, like business accounts, adopt accrual recording \citep[][par. 20.171]{esa2010european}, that is, NA \emph{``records flows at the time economic value is created, transformed, exchanged, transferred or extinguished.''} For intermediate consumption, products used in the production process are recorded and valued when they enter the process \citep[][par. 3.91]{esa2010european}. 

By contrast, the payment data show when the payment was made and received (without delays). As noted in \citet{esa2010european}, accrual basis \emph{``is different from cash recording and, in principle, from due-for-payment recording, defined as the latest time payments can be made without additional charges or penalties.''} We are also unable to identify whether payment flows refer to goods and services used in the production process at the time of the transactions. Some transactions may refer to inventories, thus contributing to the observed difference between the two data sources. 

The difference in the recording time can be important in some applications, such as real-time supply chain analyses. In many industries, suppliers are paid with a delay, and cash flow financing is an important part of credit activities with financial intermediaries specialising in supply chain financing \citep{gelsomino2016supply}. 

\paragraph{Investment in physical capital}
The intermediate use table shows the value of \textit{intermediate} goods and services exchanged between industries, that is, \emph{``goods and services consumed as inputs by a process of production, excluding fixed assets whose consumption is recorded as consumption of fixed capital. The goods and services are either transformed or used up by the production process''} \citep[][par. 3.88]{esa2010european}. 
By contrast, payments between industries are observed for multiple reasons, including payments related to capital investments and debt repayments. The latter occurs when firms finance at least part of their investment via debt and generate credit flows. 
On the other hand, the payment data potentially embodies an investment network, that could be separated by distinguishing capital and intermediate goods-producing businesses at the 5-digit SIC level. 
Such an investment network may be a valuable supplement when connecting short-term business cycles to investment dynamics and long-term growth \citep{vom2022investment}. 

\paragraph{Financial services}
In NAs, the output of financial intermediation services arises from two components \citep[p.106]{eurostat2008eurostat}: first, financial institutions receive direct fees and commissions explicitly charged. We expect to see such fees in the payment data. 
Second, NAs consider that financial intermediaries provide credit services and the value of these services can be estimated by finding the margin taken by financial institutions on the credit they make. This margin is estimated by comparing the interest rate at which banks borrow, and the one at which they make loans.  As a result, in NAs, the payment between an industry and the financial sector represents the value of financial services provided. 

In the payment data, by contrast, we observe the raw flows of funds, rather than the margin. Hence, we expect credit flows in both directions: flows of money from a creditor to a debtor, and, subsequently, reimbursements from a debtor to a creditor. In addition, in some cases, banks may act as an intermediary of intermediate trade, if businesses rely on supply chain finance services provided by the financial sector \citep{gelsomino2016supply}.
These flows can be large and would not appear in the IOTs, leading to an over-representation of the financial sector in the payment-based IOTs (see also \ref{app:scale_differences}, \ref{app:tab:top10_influence_vector}).

\paragraph{Trade and transport margins}
Within the SUTs, the supply table is valued at basic prices, while the use tables are valued at purchasers' prices. The transition from basic to purchasers' prices involves reallocating trade and transport margins. 
The output of retail and wholesale sectors equals total trade margins and is included in the supply table, while their services appear as an empty row in the intermediate use table (they are included in the purchasers' prices). As a result, when a firm from industry $i$ buys an intermediate good from industry $j$ via a wholesaler/retailer $k$, SUTs record the flow between industry $i$ and $j$ directly, adding another flow from the buyer to the wholesaler/retailer $k$ to account for the payment of trade services (part of the trade margins).

In the payment data, only direct payments are present. This causes two issues. 
First, there is a double-counting issue. In NAs, the value of goods bought for resale is counted only once, when it flows from industry $j$ to industry $i$. By contrast, the payment data likely capture both flows from the wholesaler/retailer $k$ to the seller $j$, and from the buyer $i$ to industry $k$. This means that the value of an intermediate good would appear twice in the payment data (and include trade services). 
Second, there is a misallocation issue. In the SUTs at purchasers' prices, we see a flow between industry $i$ and $j$, and no flow between the wholesaler/retailer and the supplier. By contrast, in the payment data, we would not see a flow between industries $i$ and $j$, but we would observe flows between industries $k$ and $j$ and between industries $k$ and $i$. 
Similar issues arise for transport margins. 

\paragraph{Distributive transactions} Distributive transactions are those where the value added generated by production is redistributed \citep[][par. 4.01]{esa2010european}. This includes compensation of employees, taxes on production and imports, subsidies, property income, and other current transfers. 
Within the NAs framework, such elements are outside inter-industry intermediate transaction matrices. Some flows associated with these transfers may appear in our payment data, for instance, dividends and interests, insurance premiums and settlements, or taxes and subsidies. This is important in our comparison exercise for some industries, such as public administration, which acts as a source and destination of various redistributive transactions, especially taxes and subsidies. 
Within the SUTs, the difference between taxes and subsidies is used to move from basic prices to producers’ prices. Thus, this contributes to the value of the products supplied in the economy as available in the supply table, equalling total use. By contrast, if the flow of funds captures subsidies to and taxes from businesses, we observe the flow of payments from/to public administration to other industries.    

\paragraph{International trade} In NAs, the Supply Tables show the total supply of CPA-classified products in an economy, with an extra column showing ``Imports''. Similarly, the Use Tables (``Final demand'') show domestic use of CPA-coded products and include an additional column for ``Exports'' to account for the use by non-domestic entities, ensuring total supply equals total use.
However, the final symmetric IOTs only show exports as a final demand column, and imports are integrated into the inter-industry matrix, to ensure that each column shows meaningful input requirements, based on estimates of typical industry needs, irrespective of where it sources it from. The ``Combined Use'' matrix obtained from the SUTs incorporates imports at the product-industry level.

In the payment data, we do not observe non-domestic payment flows. Transactions in the Bacs payment system can only be made in £ and require businesses to register an account in the system. Setting up an account is relatively costly and businesses need to fulfil stringent eligibility criteria or use a third-party provider. Most international transactions are made through the SWIFT payment system and, thus, are excluded from our payment data. Trade by international entities is only captured if (1) foreign entities trade with UK businesses in £, (2) have a Bacs account, and (3) are registered in Companies House, the UK business register, which is used for industry classification. These conditions may be potentially met in some cases. Further, it may be possible that the UK financial, retail and wholesale sectors serve as intermediaries for cross-national payment flows, being absorbers of exports and sources of imports. 

In summary, we may think of IOTs as including imports but not exports, while the payment data exclude the  majority of imports and exports. 

\paragraph{Unit of analysis and industrial classification}
NAs group institutional units either based on their function or kind of activity \citep[par. 1.55-1.56]{esa2010european}. Institutional units are \emph{``economic entities that are capable of owning goods and assets, of incurring liabilities and of engaging in economic activities and transactions with other units in their own right''}, and are grouped into five distinct sectors: financial and non-financial corporations, households, general government, and non-profit institutions serving households (NPISH) \citep[par 1.57]{esa2010european}. In our payment data, anonymised and aggregated Bacs transactions between industries are derived from a sample of organisations that are Bacs service users, which makes them our original unit of analysis. This includes public and private entities.

As a result of this process, two main issues arise. 
First, there is a ``headquarters effect''. Payments to/from an enterprise might be captured under the industry classification of its headquarters, although these payments might refer to subsidiaries producing other goods. 
Second, there is a risk that entities are classified into the incorrect sector (e.g., an NPISH classified as a non-financial corporation). The observed payments between industries might be affected by the different routes available to access Bacs services. Where organisations use an intermediary, the payment flows might be attributed to the intermediaries rather than directly between the organisations paying or receiving funds \citep{ons2023interindustrymethods}. 
Analytical IOTs are built from surveys that attempt to consider the multi-product nature of firms. Here, instead, entities are classified into a single industry. 

Further issues arise with the classification of activities, particularly public services. The classification in the payment data are derived from matching Bacs service users to SIC codes indicated in the UK business register Companies House. 
While this may be consistent with the NA data collected through business surveys, NAs follow different principles for certain sectors, especially public administration, defence and compulsory social security services (084). The O84-activities are defined as being of a governmental nature provided \emph{``as non-market services and valued accordingly''} \citep[][par. 3.84]{esa2010european}. In the payment data, public administration entities are identified by a ``hard coding'' approach, using ONS-internal data to inform the account classification. The payment data also include transactions of private companies that use SIC codes related to the public administration to describe the nature of their business during the Companies House registration. Such codes include, for example, consultancy and technical services for defence, fire protection, or support for international trade. 

\paragraph{Informal sector} NAs should, in principle, estimate the output and income from the informal sector. It is not necessarily clear whether this output appears in the payment data. Such transactions would likely be made with cash or card payments rather than electronic transactions. To the extent that informal activities are accurately represented in IOTs, and are absent from data based on electronic payments, we expect industries with high informal activities to be under-represented in payment data, compared to NAs. However, neither of the assumptions can be verified or appear plausible a priori.

\FloatBarrier
\section{Stylised facts of the granular data}
\label{sec:stylised_facts_5digit}
Now, we study the most granular 5-digit data. A direct comparison to official statistics is infeasible, as such granular data does not (yet) exist at a macroeconomic scale.\footnote{An exception are the granular IOTs available for the US, which, however, are only available at a quinquennial basis \citep{hotte2023demand}.} 
To benchmark the data, we evaluate the data by its ability to reproduce two stylised facts documented in the literature on economic networks: 
\begin{enumerate}
    \item The average correlation of growth rates at a given network distance apart decreases with network distance (Sec. \ref{sec:corr_growth_rates_main}).
    \item The CCDF of the so-called Katz-Bonacich centrality exhibits a power law-like behaviour with a tail exponent $1< \gamma <2$ (Sec. \ref{sec:influence_vector}), implying that shocks at the industry level can lead to aggregate fluctuations, for example, in GDP.
\end{enumerate}
Consistency with these stylised facts suggests that the granular payment data are valuable for economic network research. 
\FloatBarrier
\subsection{Correlation of growth rates}
\label{sec:corr_growth_rates_main}
\citet{carvalho2014from, mungo2023revealing} documented for industry- and firm-level data that the correlation of rates between a pair of industries (firms) decreases with their distance in the network, where the distance refers to the shortest path of input linkages in the network that connects the two sectors.  
Here, we test whether this holds in the granular 5-digit network data of 601 distinct industries, and correlate industry-level growth rates of the selling and buying industry, whereby growth rates are given by the change in industry-level outputs (inputs) from a given month to the month in the subsequent year. We calculate correlations for input and output growth using count and value data, and plot the correlations against the distance. The network distances are obtained from annual aggregate input networks for the corresponding year. The colours indicate different truncation thresholds imposed on the network to remove noisy links.\footnote{The truncation procedure is the same as discussed before (Sec. \ref{subsec:aggregate_network}).} 

\begin{figure}[!h]
    \centering
    
    \caption{Correlations of growth rates}
    \label{fig:growth_correl_by_distance}
    
    \includegraphics[width=\textwidth]{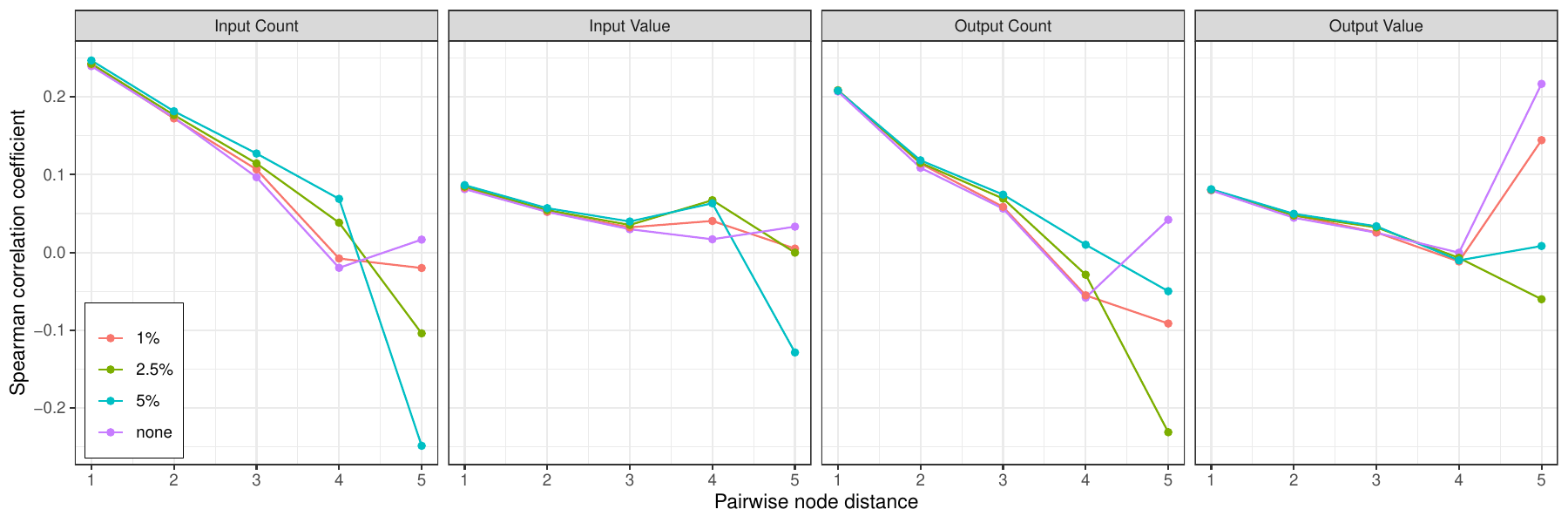}

     \justifying \scriptsize
    \noindent
    Notes: These figures illustrate the Spearman correlation coefficients between monthly (year on year) growth rates of directly and indirectly connected pairs of industries, using data from 2016 to 2019, excluding the Covid-19 period. 
    The x-axis shows the distance of the industry pairs in annual network aggregates.\footnote{The distances are the shortest path (lowest number of steps) that connects the pair of industries in the network. A value of one indicates a direct link (one step) between the pair.} 
    The colours indicate truncation thresholds imposed on the network before calculating the distances. Links with a weight (input share) below the threshold are removed (see also Sec. \ref{subsec:aggregate_network}).  

  \label{fig:corr_network_2019}
\end{figure}

Fig. \ref{fig:growth_correl_by_distance} illustrates the results, with the vertical axis showing the Spearman correlation coefficient and the horizontal axis showing different distance levels. We use the Spearman correlation due to its lower sensitivity against outliers compared to Pearson correlations used above (Sec. \ref{sec:benchmarking}). The results generally confirm that growth rate correlations decrease in the network distance. The results become noisy and even negative at large distances, which is not surprising given the sparsity and incomplete coverage of the data. 
The results are consistent across the different data types (inputs, outputs, counts, values) with steeper curves for count data.

\FloatBarrier
\subsection{Centrality distribution}
\label{sec:influence_vector}

Previous research has shown that the impact of firm- and industry-level and industry-level shocks on aggregate economic fluctuations depends on the firm or industry's network position \citep[e.g.][]{acemoglu2012network,carvalho2014from, roson2016input}. 
Negative and positive shocks occurring in an industry that plays a central role in the supply and demand linkages network tend to have larger spillover effects on other industries. 
Such supply chain spillovers are a key reason for studying economic networks. 

The ``right'' way of measuring the centrality of an industry depends on the assumption of the underlying model to study aggregate volatility and the nature of available data. For some established centrality metrics, one needs to know the whole IOT, including value-added and final demand components next to intermediate trade as captured by our payment data. 

Because we do not have final demand and value-added equivalents in our data, we compute a centrality metric that can be computed solely from the industry-industry flows. We use the \textit{influence vector}, also known as Katz-Bonacich centrality, which quantifies the impact of industry-level productivity shocks in a standard equilibrium input-output analysis with Cobb-Douglas production functions, no capital, and uniform final demand shares \citep{acemoglu2012network,magerman2016heterogeneous}. It is given by
\begin{align}
    v\equiv \frac{\alpha_L}{n}\left[\mathbf{I} - (1-\alpha_L)\textbf{W}'\right]^{-1}\mathbf{1},
    \label{eq:Influence_Vector}
\end{align}
where $\alpha_L\in (0,1]$ is the labour share of gross output, $n$ is the number of industries, $\mathbf{I}$ is an identity matrix, $\mathbf{1}$ is a vector of ones and $\textbf{W}'$ is the (column-stochastic) matrix of input shares, $\omega^{in,\alpha}_{ij}$ computed according to Eq.~\eqref{eq:shares_in_out}. The influence vector $v$ is a micro-level measure of the importance of a certain industry in the production network. An interesting theoretical result \citep{acemoglu2012network} is that its distribution is proportional to aggregate fluctuations as
\begin{align}
    \text{std}(\text{log}(\text{GDP}))\sim n^{-(1-1/\gamma)},
\end{align}
where $1<\gamma \leq 2$ is the power law exponent of the distribution of the influence vector, and $n$ is the number of firms or industries. In other words, if centralities are highly unequally distributed, micro-shocks to industries do not average out in the aggregate. 

Previous studies have measured $\gamma$ on existing input-output data, providing us with benchmark results to compare our data with. As a first step, we analyse whether the influence vector in the payment data follows a power law. 

\begin{figure}[!h]
    \centering
    \caption{CCDF of the Katz-Bonacich centrality}
    \label{fig:influence_vector}
    \includegraphics[width=0.8\textwidth]{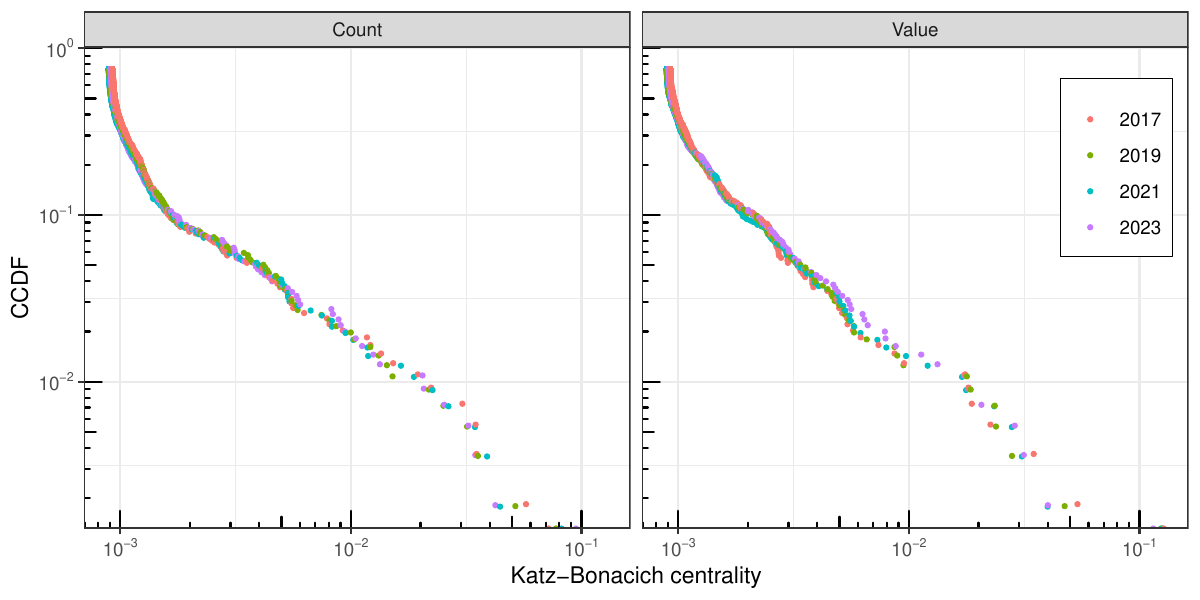}

  \justifying \scriptsize
    \noindent
    Notes: These figures illustrate the CCDF of the Katz-Bonacich centrality (see Eq.~\eqref{eq:Influence_Vector}) for different years, using a labour share parameter of $\alpha_L = 0.5$ \citep{magerman2016heterogeneous} and payment-based input share matrices based on counts and values. 
    
  \label{fig:ccdf_influence_vector}
\end{figure}

Fig. \ref{fig:ccdf_influence_vector} shows a complementary cumulative distribution (CCDF) at a log-log scale, for the input share networks from different years. The CCDF supports the idea of a heavy-tailed distribution, with most sectors scoring at low values and some sectors being extremely central. In \ref{app:tab:top10_influence_vector}, we list the top-10 sectors scoring extremely high, and find public administration to take the top rank, consistently across years and datasets, and the other ranks being taken mainly by retail and finance, and at lower levels transport and electricity. The count data generally appears to exhibit a slightly more equal distribution, with less extreme deviations among sectors. 

The close-to-linear shape of the tail of the log-log CCDF indicates a power law.
To test whether the data can be well-fitted by a power law distribution and to obtain the tail exponent $\gamma$, we use a Hill estimator \citep{Clauset_2009}. Results and test statistics of this fitting exercise are provided in \ref{app:subsec:influence_power_law}.
We find tail exponents, ranging between 1.34 and 1.69 for the value and 1.98 and 2.21 for the count data. The values for the value data are similar to those reported in the literature using firm- and industry-level data. For example, \citet{carvalho2014from} reported $\gamma=1.44$ for industry-level US data, and \citet{magerman2016heterogeneous} and \citet{bacilieri2022firm} found $\gamma \in [1.12,1.44]$, using firm-level VAT data from Belgium, Ecuador, and Hungary.\footnote{VAT data reports supplier-customer relationships amongst firms within a country. Just like our payment data, they represent flows of money and usually record transactions above a certain threshold. For instance, for Belgium, the threshold is 250€. For more details on the description of the VAT datasets from Ecuador and Hungary see \citet{bacilieri2022firm}.}
However, the significance tests suggest that the power law hypothesis can be only supported for some years, but more often when using count data. 

\FloatBarrier
\subsection{Discussion}
In the previous two subsections, we made connections to two stylised facts in the literature on economic networks:
\begin{itemize}
    \item The average correlation of growth rates of industries at a given network distance apart decreases with distance.
    \item The CCDF of the Katz-Bonacich centrality has a power law tail of $1<\gamma < 2$. 
\end{itemize}
Our results confirm an alignment with the literature \citep{bacilieri2022firm, magerman2016heterogeneous}, suggesting that the network structure of the granular payment data resemble the structure of other large-scale economic networks, that have been successfully used in economic research. 

The decrease in the correlation of growth rates with the network distance indicates network effects: industries grow when their neighbours (suppliers and customers) grow. This can be informative for clusters of industrial growth, and cross-industrial spillover effects \citep{hotte2023demand, carvalho2014from, roson2016input}, telling policymakers about which industries to nurture to promote growth in particular sectors or regions \citep{luo2013industries, oosterhaven2021interregional, dietzenbacher2002interregional, kitsos2023industrial}. 

Further, we obtain values around 1.5 for the tail exponents of the CCDF of the Katz-Bonacich centrality in agreement with exponents found in previous studies. Tail-exponents of $1<\gamma<2$ indicate that micro-level fluctuations can be drivers of large aggregate fluctuations, suggesting a need to monitor the economy at a granular level to understand aggregate outcomes.

The analysis of centrality has also revealed some ``biases'' towards the public sector, finance, and retail compared to the NA perspective (see Sec. \ref{sec:conceptual}). Some of them are also present in other firm-level data sets \citep[see][]{bacilieri2022firm}. This does not hamper the usefulness of this data for research but may affect the validity of assumptions made in theoretical and empirical models applied to the data. 

\section{Conclusion and outlook}
\label{sec:conclusion}
This study offers a first-of-its-kind economic validation of a high-frequency, granular dataset on inter-industry financial transactions in the UK. By capturing realised monetary flows with unprecedented detail, this data source opens new frontiers in economic analysis, national statistics, and policy evaluation. Its monthly resolution enables real-time tracking of business dynamics, yet its integration with conventional NA frameworks remains a challenge due to the absence of standardised reporting structures.

Our benchmarking exercises reveal strong correlations between aggregate payment values and monetary aggregates, while transaction counts serve as a distinct indicator of real economic activity. The link between payment counts and real GDP suggests that this data could serve as a novel proxy for business dynamism by tracking deviations from routine financial commitments such as fees, royalties, and loan repayments. While count-based metrics have been rarely used in economic research, our analysis suggests their potential to differentiate real from nominal fluctuations.

At the network level, our payment data exhibit lower density than traditional IOTs, yet retains structural consistency when major linkages are considered. However, transaction-level discrepancies highlight conceptual differences related to industry classifications, the treatment of intermediary industries such as retail and finance, and the time of recording. These must be addressed in future applications in economic modelling.

Despite these challenges, raw payment data that become increasingly available beyond the UK provide a complementary lens on inter-industrial trade, illuminating realised monetary flows often obscured by national accounting conventions \citep[cf.][]{simon1995organizations}. This is particularly valuable for studying industrial diversification, technological adaptation, and policy-induced shifts that may not be readily captured in official statistics. Our validation exercises at the 5-digit SIC level further reinforce the credibility of this dataset as a robust input for economic network analysis at a more granular level.

By mapping out the challenges and opportunities of this innovative data source, we aim to pave the way for cutting-edge research in economic systems. Future applications include real-time economic nowcasting \citep{mantziou2024gdp}, granular production network dynamics, and early-warning indicators for supply chain disruptions. The potential to derive regional breakdowns also opens avenues for studying Brexit’s economic impact, supply chain resilience, and policy responses for levelling up and net-zero transitions.

This work relied on an experimental version of the data \citep[see also][]{ons2023interindustry, ons2023interindustrymethods}. As this dataset continues to evolve, with improvements in coverage, classification, and timely availability, it holds promise as a transformative tool for economic measurement. We hope this study serves as a primer and guide for its future integration into empirical research and policy frameworks.

\newpage
\renewcommand \bibname{References}
\printbibliography

\newpage
\appendix
\renewcommand{\appendixname}{Appendix}
\renewcommand{\thesection}{\Alph{section}} \setcounter{section}{0}
\renewcommand{\thefigure}{\Alph{section}.\arabic{figure}} \setcounter{figure}{0}
\renewcommand{\thetable}{\Alph{section}.\arabic{table}} \setcounter{table}{0}
\renewcommand{\theequation}{\Alph{section}.\arabic{table}} \setcounter{equation}{0}

\FloatBarrier
\section{Payment systems background}

\subsection{Basic concepts}
\label{app:subsec:paysystems_basics}
Payments are made in different ways, for example, using cash, credit and debit cards, bank transfers, mobile payments, or cheques, and thereby rely on various interconnected \emph{payment systems}. 
In this work, we focus on anonymised and aggregated electronic payments in £ extracted from the \emph{infrastructure} of the Bankers' Automated Clearing System (Bacs), which is one key system used by UK businesses for bank transfers. Infrastructure data differs from transaction data obtained from single banks \citep{buda2023national, carvalho2021tracking, ialongo2022reconstructing}, as the infrastructure connects different banks and other payment service providers (PSPs) to transfer funds between the accounts of their clients. 

Different payment infrastructures co-exist and are dependent on the \emph{payment scheme}. Loosely speaking, a payment scheme is a set of rules on how to transfer funds between accounts at different PSPs. The rules cover, for example, transaction speed and limit, and the definition of payment instruments, such as direct debits or credit transfers. PSPs can decide whether they join a scheme, but usually, all major PSP within an economic area usually use the same schemes.

Transferring funds involves two steps: \emph{clearing} and \emph{settlement}. Roughly speaking, clearing is the exchange of messages about an obligation to be established, sometimes including an inquiry of whether funds on the payer's account are sufficient. Settlement is the realisation of the transfer, which often happens with a time delay in pre-determined settlement cycles and on a net basis. 
Net settlement means that PSPs only transfer the net of their mutual obligations arising from multiple transactions made within the cycle \citep{bis2016glossary}. Unlike other work using payment system data provided by central banks \citep{aprigliano2019using}, our data is collected at the clearing level. This preserves the account-level network structure among businesses, which is otherwise hidden by financial intermediaries. 

One can distinguish \emph{wholesale} and \emph{retail} payment systems, whereby wholesale is mostly used for high-value transactions settled in real-time gross settlement. Retail payment systems often settle on a net basis and are primarily used in everyday economic activity \citep{aprigliano2019using}. Bacs is one of the major retail payment systems in the UK. The UK's wholesale payment system is CHAPS, operated by the Bank of England (BoE). Non-financial businesses rarely use it for everyday transactions as it is costly, although they may still choose it for high-value and time-sensitive payments. 

 \subsection{Innovation and change in payments}
\label{app:subsec:innovation_payments_explained}

One key challenge for using payment data in research is their responsiveness to crises, regulation, attempts for international harmonisation, and innovation. This can affect businesses' and consumers' choice of how to make payments, as exemplified by the decline of cash and cheques, and the rise of card and FPS payments \citep{UKfinance2022UKpayments, bodley2022fintech, jackson2018innovation}.
Until now, most innovations have been limited to the relationship between PSPs and their customers, such as new payment instruments and services, connecting services, or user interfaces. These innovations were driven mainly by digitalisation and enabled by regulation after the financial crisis. This was aligned with high-level operational changes in the UK payment system, such as the introduction of FPS.

Since 2015, there has been an ongoing transformation that will likely affect all major schemes operated by Pay.UK. The Payment System Regulator (PSR) outlined a strategy to build a ``new payments architecture'' (NPA) \citep{PSO2017report}. One of the goals of the NPA is the replacement of the existing retail payment systems (Bacs, FPS, ICS) by a uniform scheme and infrastructure, providing a comprehensive technical update, and a higher compatibility with digitalisation, new consumer habits, and international developments \citep{bodley2022fintech}. 
So far, these plans have not yet been realised, and the impact on payment data is complex to evaluate ex-ante. In the best case, harmonising payments under a uniform architecture would improve the coverage, assuming that matching accounts with businesses and industries would still be possible.

\FloatBarrier

\section{Concordance table}
\label{app:concordance}
Table \ref{tab:concordance} shows how industries classified by 5-digit SIC codes are re-allocated to CPA codes used in the official ONS IOT and NA data \citep{ons2009sic, eurostat2015cpa}. The 5-digit SIC codes are more disaggregate and aggregated into 105 CPA classes. The codes in the first column (SIC) are short for the first 2-4 digits of the 5-digit codes. All industries with these digits as leading digits are aggregated into the respective CPA category. The ``$\cdot$''s in the columns of the table indicate which SIC codes belong to a more aggregate CPA category. 

\begingroup\tiny \centering
\begin{longtable}{|p{0.5cm}|p{6cm}|p{1cm}|p{6cm}|}
  \hline \\[-1.8ex] 
SIC & SIC names & CPA & CPA names \\[1ex]
  \hline  \\[-1.8ex] 
  
  \endhead
01 & Crop and animal production, hunting and related service activities & A01 & Products of agriculture, hunting and related services \\ 
  02 & Forestry and logging & A02 & Products of forestry, logging and related services \\ 
  03 & Fishing and aquaculture & A03 & Fish and other fishing products; aquaculture products; support services to fishing \\ 
  05 & Mining of coal and lignite & B05 & Coal and lignite \\ 
  06 & Extraction of crude petroleum and natural gas & B06-F7 & Extraction Of Crude Petroleum And Natural Gas \& Mining Of Metal Ores \\ 
  07 & Mining of metal ores &$\cdot$&$\cdot$\\ 
  08 & Other mining and quarrying & B08 & Other mining and quarrying products \\ 
  09 & Mining support service activities & B09 & Mining support services \\ 
  101 & Preserved meat and meat products & C101 & Preserved meat and meat products \\ 
  102 & Processing and preserving of fish, crustaceans and molluscs & C102-3 & Processed and preserved fish, crustaceans, molluscs, fruit and vegetables \\ 
  103 & Processing and preserving of fruit and vegetables &$\cdot$&$\cdot$\\ 
  104 & Vegetable and animal oils and fats & C104 & Vegetable and animal oils and fats \\ 
  105 & Dairy products & C105 & Dairy products \\ 
  106 & Grain mill products, starches and starch products & C106 & Grain mill products, starches and starch products \\ 
  107 & Bakery and farinaceous products & C107 & Bakery and farinaceous products \\ 
  108 & Other food products & C108 & Other food products \\ 
  109 & Prepared animal feeds & C109 & Prepared animal feeds \\ 
  1101 & Distilling, rectifying and blending of spirits & C11.01-6 \& C12 & Alcoholic beverages \& Tobacco products \\ 
  1102 & Manufacture of wine from grape &$\cdot$&$\cdot$\\ 
  1103 & Manufacture of cider and other fruit wines &$\cdot$&$\cdot$\\ 
  1104 & Manufacture of other non-distilled fermented beverages &$\cdot$&$\cdot$\\ 
  1105 & Manufacture of beer &$\cdot$&$\cdot$\\ 
  1106 & Manufacture of malt &$\cdot$&$\cdot$\\ 
  1107 & Manufacture of soft drinks & C1107 & Soft drinks \\ 
  12 & Manufacture of tobacco products &$\cdot$&$\cdot$\\ 
  13 & Manufacture of textiles & C13 & Textiles \\ 
  14 & Manufacture of wearing apparel & C14 & Wearing apparel \\ 
  15 & Manufacture of leather and related products & C15 & Leather and related products \\ 
  16 & Manufacture of wood and of products of wood and cork, except furniture; manufacture of articles of straw and plaiting materials & C16 & Wood and of products of wood and cork, except furniture; articles of straw and plaiting materials \\ 
  17 & Manufacture of paper and paper products & C17 & Paper and paper products \\ 
  18 & Printing and reproduction of recorded media & C18 & Printing and recording services \\ 
  19 & Manufacture of coke and refined petroleum products & C19 & Coke and refined petroleum products \\ 
  2011 & Manufacture of industrial gases & C20A & Industrial gases, inorganics and fertilisers (all inorganic chemicals) - 20.11/13/15 \\ 
  2012 & Manufacture of dyes and pigments & C20C & Dyestuffs, agro-chemicals - 20.12/20 \\ 
  2013 & Manufacture of other inorganic basic chemicals &$\cdot$&$\cdot$\\ 
  2014 & Manufacture of other organic basic chemicals & C20B & Petrochemicals - 20.14/16/17/60 \\ 
  2015 & Manufacture of fertilisers and nitrogen compounds &$\cdot$&$\cdot$\\ 
  2016 & Manufacture of plastics in primary forms &$\cdot$&$\cdot$\\ 
  2017 & Manufacture of synthetic rubber in primary forms &$\cdot$&$\cdot$\\ 
  2020 & Manufacture of pesticides and other agrochemical products &$\cdot$&$\cdot$\\ 
  203 & Paints, varnishes and similar coatings, printing ink and mastics & C203 & Paints, varnishes and similar coatings, printing ink and mastics \\ 
  204 & Soap and detergents, cleaning and polishing preparations, perfumes and toilet preparations & C204 & Soap and detergents, cleaning and polishing preparations, perfumes and toilet preparations \\ 
  205 & Other chemical products & C205 & Other chemical products \\ 
  2060 & Manufacture of man-made fibres &$\cdot$&$\cdot$\\ 
  21 & Manufacture of basic pharmaceutical products and pharmaceutical preparations & C21 & Basic pharmaceutical products and pharmaceutical preparations \\ 
  22 & Manufacture of rubber and plastic products & C22 & Rubber and plastic products \\ 
  231 & Manufacture of glass and glass products & C23 other & Glass, refractory, clay, other porcelain and ceramic, stone and abrasive products - 23.1-4/7-9 \\ 
  232 & Manufacture of refractory products &$\cdot$&$\cdot$\\ 
  233 & Manufacture of clay building materials &$\cdot$&$\cdot$\\ 
  234 & Manufacture of other porcelain and ceramic products &$\cdot$&$\cdot$\\ 
  235 & Manufacture of cement, lime and plaster & C235-6 & Cement, lime, plaster and articles of concrete, cement and plaster \\ 
  236 & Manufacture of articles of concrete, cement and plaster &$\cdot$&$\cdot$\\ 
  237 & Cutting, shaping and finishing of stone &$\cdot$&$\cdot$\\ 
  239 & Manufacture of abrasive products and non-metallic mineral products n.e.c. &$\cdot$&$\cdot$\\ 
  241 & Manufacture of basic iron and steel and of ferro-alloys & C241-3 & Basic iron and steel \\ 
  242 & Manufacture of tubes, pipes, hollow profiles and related fittings, of steel &$\cdot$&$\cdot$\\ 
  243 & Manufacture of other products of first processing of steel &$\cdot$&$\cdot$\\ 
  244 & Manufacture of basic precious and other non-ferrous metals & C244-5 & Other basic metals and casting \\ 
  245 & Casting of metals &$\cdot$&$\cdot$\\ 
  251 & Manufacture of structural metal products & C25 other & Fabricated metal products, excl. machinery and equipment and weapons \& ammunition - 25.1-3/25.5-9 \\ 
  252 & Manufacture of tanks, reservoirs and containers of metal &$\cdot$&$\cdot$\\ 
  253 & Manufacture of steam generators, except central heating hot water boilers &$\cdot$&$\cdot$\\ 
  254 & Weapons and ammunition & C254 & Weapons and ammunition \\ 
  255 & Forging, pressing, stamping and roll-forming of metal; powder metallurgy &$\cdot$&$\cdot$\\ 
  256 & Treatment and coating of metals; machining &$\cdot$&$\cdot$\\ 
  257 & Manufacture of cutlery, tools and general hardware &$\cdot$&$\cdot$\\ 
  259 & Manufacture of other fabricated metal products &$\cdot$&$\cdot$\\ 
  26 & Manufacture of computer, electronic and optical products & C26 & Computer, electronic and optical products \\ 
  27 & Manufacture of electrical equipment & C27 & Electrical equipment \\ 
  28 & Manufacture of machinery and equipment n.e.c. & C28 & Machinery and equipment n.e.c. \\ 
  29 & Manufacture of motor vehicles, trailers and semi-trailers & C29 & Motor vehicles, trailers and semi-trailers \\ 
  301 & Ships and boats & C301 & Ships and boats \\ 
  302 & Manufacture of railway locomotives and rolling stock & C30 other & Other transport equipment - 30.2/4/9 \\ 
  303 & Air and spacecraft and related machinery & C303 & Air and spacecraft and related machinery \\ 
  304 & Manufacture of military fighting vehicles &$\cdot$&$\cdot$\\ 
  309 & Manufacture of transport equipment n.e.c. &$\cdot$&$\cdot$\\ 
  31 & Manufacture of furniture & C31 & Furniture \\ 
  32 & Other manufacturing & C32 & Other manufactured goods \\ 
  3311 & Repair of fabricated metal products & C33 other & Rest of repair; Installation - 33.11-14/17/19/20 \\ 
  3312 & Repair of machinery &$\cdot$&$\cdot$\\ 
  3313 & Repair of electronic and optical equipment &$\cdot$&$\cdot$\\ 
  3314 & Repair of electrical equipment &$\cdot$&$\cdot$\\ 
  3315 &$\cdot$& C3315 & Repair and maintenance of ships and boats \\ 
  3316 &$\cdot$& C3316 & Repair and maintenance of aircraft and spacecraft \\ 
  3317 & Repair and maintenance of other transport equipment &$\cdot$&$\cdot$\\ 
  3319 & Repair of other equipment &$\cdot$&$\cdot$\\ 
  332 & Installation of industrial machinery and equipment &$\cdot$&$\cdot$\\ 
  351 & Electricity, transmission and distribution & D351 & Electricity, transmission and distribution \\ 
  352 & Manufacture of gas; distribution of gaseous fuels through mains & D352-3 & Gas; distribution of gaseous fuels through mains; steam and air conditioning supply \\ 
  353 & Steam and air conditioning supply &$\cdot$&$\cdot$\\ 
  36 & Water collection, treatment and supply & E36 & Natural water; water treatment and supply services \\ 
  37 & Sewerage & E37 & Sewerage services; sewage sludge \\ 
  38 & Waste collection, treatment and disposal activities; materials recovery & E38 & Waste collection, treatment and disposal services; materials recovery services \\ 
  39 & Remediation activities and other waste management services. & E39 & Remediation services and other waste management services \\ 
  41 & Construction of buildings & F41-43 & Construction \\ 
  42 & Civil engineering &$\cdot$&$\cdot$\\ 
  43 & Specialised construction activities &$\cdot$&$\cdot$\\ 
  45 & Wholesale and retail trade and repair of motor vehicles and motorcycles & G45 & Wholesale and retail trade and repair services of motor vehicles and motorcycles \\ 
  46 & Wholesale trade, except of motor vehicles and motorcycles & G46 & Wholesale trade services, except of motor vehicles and motorcycles \\ 
  47 & Retail trade, except of motor vehicles and motorcycles & G47 & Retail trade services, except of motor vehicles and motorcycles \\ 
  491 & Passenger rail transport, interurban & H491-2 & Rail transport services \\ 
  492 & Freight rail transport &$\cdot$&$\cdot$\\ 
  493 & Other passenger land transport & H493-5 & Land transport services and transport services via pipelines, excluding rail transport \\ 
  494 & Freight transport by road and removal services &$\cdot$&$\cdot$\\ 
  495 & Transport via pipeline &$\cdot$&$\cdot$\\ 
  50 & Water transport & H50 & Water transport services \\ 
  51 & Air transport & H51 & Air transport services \\ 
  52 & Warehousing and support activities for transportation & H52 & Warehousing and support services for transportation \\ 
  53 & Postal and courier activities & H53 & Postal and courier services \\ 
  55 & Accommodation & I55 & Accommodation services \\ 
  56 & Food and beverage service activities & I56 & Food and beverage serving services \\ 
  58 & Publishing activities & J58 & Publishing services \\ 
  59 & Motion picture, video and television programme production, sound recording and music publishing activities & J59-60 & Motion Picture, Video \& TV Programme Production, Sound Recording \& Music Publishing Activities \& Programming And Broadcasting Activities \\ 
  60 & Programming and broadcasting activities &$\cdot$&$\cdot$\\ 
  61 & Telecommunications & J61 & Telecommunications services \\ 
  62 & Computer programming, consultancy and related activities & J62 & Computer programming, consultancy and related services \\ 
  63 & Information service activities & J63 & Information services \\ 
  64 & Financial service activities, except insurance and pension funding & K64 & Financial services, except insurance and pension funding \\ 
  651 & Insurance & K65.1-3 & Insurance, reinsurance and pension funding services, except compulsory social security \\ 
  652 & Reinsurance &$\cdot$& $\cdot$ \\ 
  653 & Pension funding &$\cdot$&$\cdot$\\ 
  66 & Activities auxiliary to financial services and insurance activities & K66 & Services auxiliary to financial services and insurance services \\ 
  681 & Buying and selling of own real estate & L68 BX L683 & Real estate services, excluding on a fee or contract basis and imputed rent \\ 
  682 & Owner-Occupiers' Housing Services & L68A & Owner-Occupiers' Housing Services \\ 
 $\cdot$& Renting and operating of own or leased real estate &$\cdot$&$\cdot$\\ 
  683 & Real estate services on a fee or contract basis & L683 & Real estate services on a fee or contract basis \\ 
  691 & Legal services & M691 & Legal services \\ 
  692 & Accounting, bookkeeping and auditing services; tax consulting services & M692 & Accounting, bookkeeping and auditing services; tax consulting services \\ 
  70 & Activities of head offices; management consultancy activities & M70 & Services of head offices; management consulting services \\ 
  71 & Architectural and engineering activities; technical testing and analysis & M71 & Architectural and engineering services; technical testing and analysis services \\ 
  72 & Scientific research and development & M72 & Scientific research and development services \\ 
  73 & Advertising and market research & M73 & Advertising and market research services \\ 
  74 & Other professional, scientific and technical activities & M74 & Other professional, scientific and technical services \\ 
  75 & Veterinary activities & M75 & Veterinary services \\ 
  77 & Rental and leasing activities & N77 & Rental and leasing services \\ 
  78 & Employment activities & N78 & Employment services \\ 
  79 & Travel agency, tour operator and other reservation service and related activities & N79 & Travel agency, tour operator and other reservation services and related services \\ 
  80 & Security and investigation activities & N80 & Security and investigation services \\ 
  81 & Services to buildings and landscape activities & N81 & Services to buildings and landscape \\ 
  82 & Office administrative, office support and other business support activities & N82 & Office administrative, office support and other business support services \\ 
  84 & Public administration and defence; compulsory social security & O84 & Public administration and defence services; compulsory social security services \\ 
  85 & Education & P85 & Education services \\ 
  86 & Human health activities & Q86 & Human health services \\ 
  87 & Residential care activities & Q87-88 & Residential Care \& Social Work Activities \\ 
  88 & Social work activities without accommodation &$\cdot$&$\cdot$\\ 
  90 & Creative, arts and entertainment activities & R90 & Creative, arts and entertainment services \\ 
  91 & Libraries, archives, museums and other cultural activities & R91 & Libraries, archives, museums and other cultural services \\ 
  92 & Gambling and betting activities & R92 & Gambling and betting services \\ 
  93 & Sports activities and amusement and recreation activities & R93 & Sports services and amusement and recreation services \\ 
  94 & Activities of membership organisations & S94 & Services furnished by membership organisations \\ 
  95 & Repair of computers and personal and household goods & S95 & Repair services of computers and personal and household goods \\ 
  96 & Other personal service activities & S96 & Other personal services \\ 
  97 & Activities of households as employers of domestic personnel & T97 & Services of households as employers of domestic personnel \\ 
   \hline
\hline
\caption{Concordance table from SIC to CPA codes.} 
\label{tab:concordance}
\end{longtable}
\endgroup

\FloatBarrier

\section{Additional material: comparison to existing data}
\subsection{Direct debits and credits}
\label{app:sec:macro_benchmarking}
Bacs Direct Debits (DD) and Direct Credit (DC) contain different kinds of economic information. 

 \begin{figure}[!h]
    \centering   
    
    \caption{Monthly payment data, direct debits and direct credits}
    \label{app:fig:ts_Bacs_instruments_benchmarking}

    \begin{subfigure}[b]{\textwidth}
        \centering
        \includegraphics[width=\textwidth]{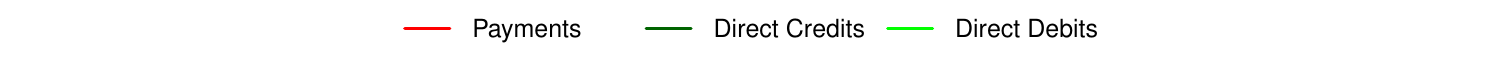}        
    \end{subfigure}
    
    \vspace{-0.225cm}

    \begin{subfigure}[b]{0.49\textwidth}
        \centering 
        \includegraphics[width=0.8\textwidth]{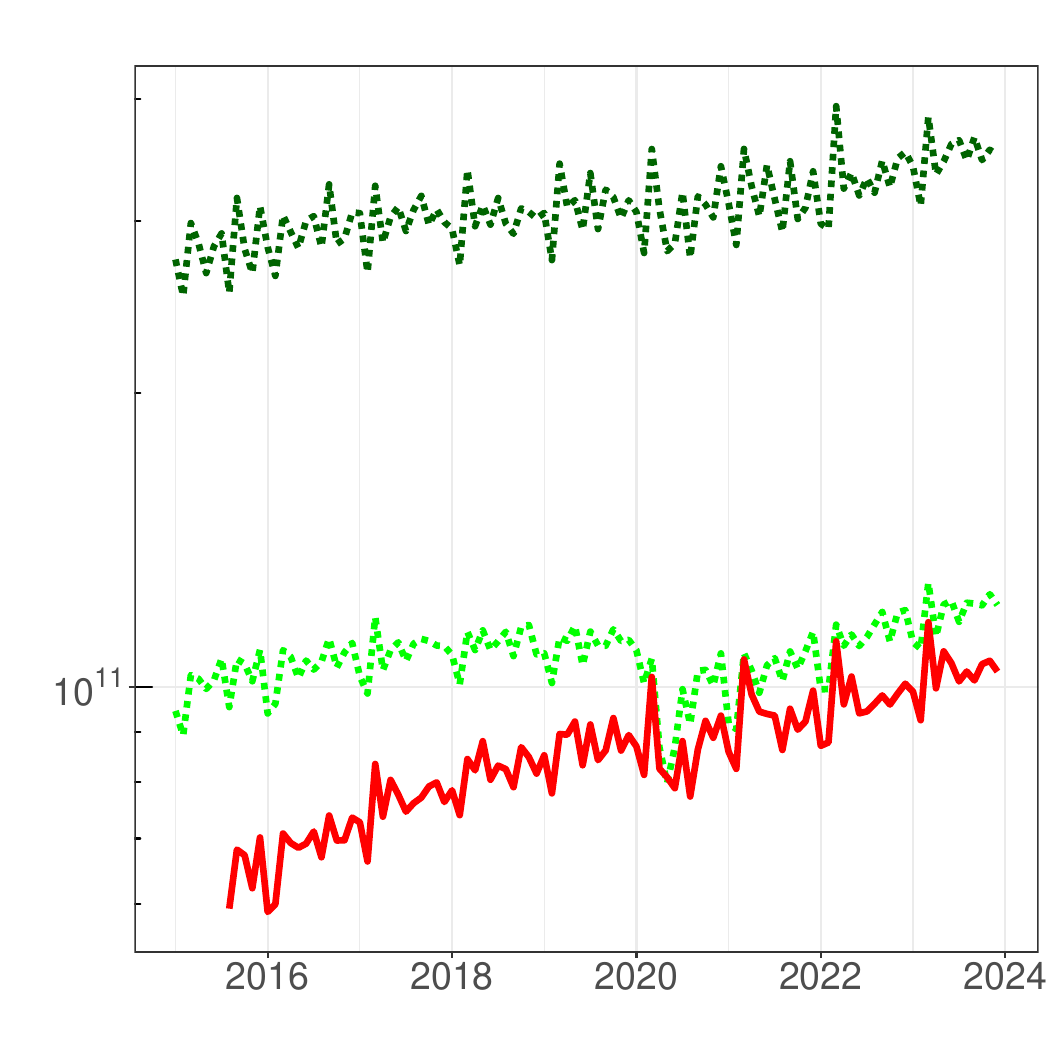}
        \caption{Values}
        \label{app:fig:ts_bacs_values}
    \end{subfigure}
    \begin{subfigure}[b]{0.49\textwidth}
        \centering
        \includegraphics[width=0.8\textwidth]{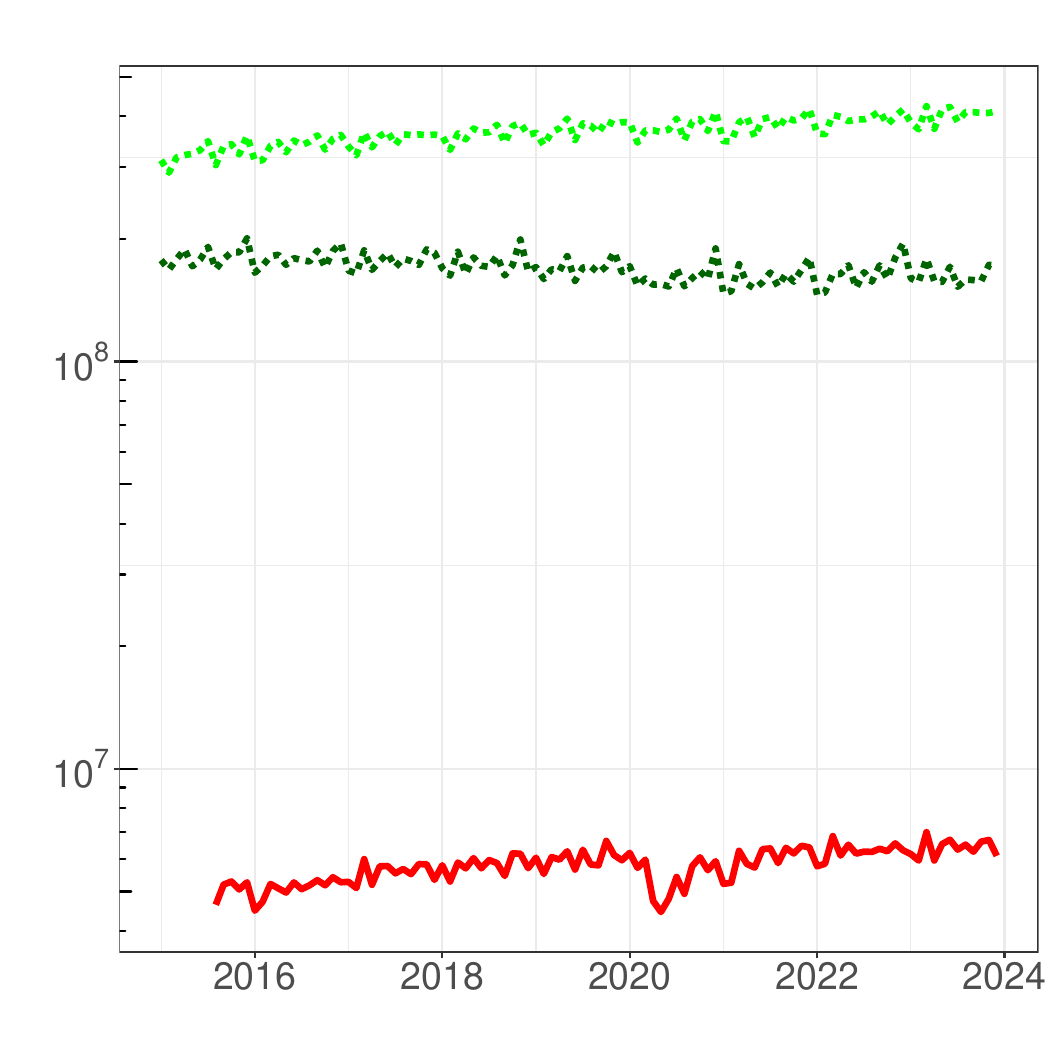}
        \caption{Counts}
        \label{app:fig:ts_bacs_counts}
    \end{subfigure}    

    \justifying \scriptsize \noindent
    Notes: The vertical axis is scaled at a log-10 scale. Payments (red) are monthly aggregates of our data. Bacs Direct Debit and Direct Credit data is downloaded from \citet{payuk2023historicaldata}. 

\end{figure}

Fig. \ref{app:fig:ts_Bacs_instruments_benchmarking} shows a time series of monthly data of the payment data and Bacs DB and DC by nominal values in log £ millions (Fig. \ref{app:fig:ts_bacs_values}) and counts in log £ 1,000s (Fig. \ref{app:fig:ts_bacs_counts}). 
The time series indicates a persistent rise in transaction values for DC and our payment data. Bacs DB exhibit a steep downward kink during the first Covid-19 lock down and a monotonous recovery thereafter, back to the pre-Covid level. Compared to DC and the payment data, the overall growth in DB values from 2015 to 2023 was modest. 
Since the start of the Covid-19 pandemic, the aggregate value of the payment data has risen to almost the same aggregate value than DDs. 
DCs are the largest share of values processed through Bacs, despite corresponding to a smaller share of counts compared to DD. As discussed briefly in \ref{app:sec:macro_benchmarking}, Bacs DB and DC tend to differ by patterns over time, responses to Covid-19, and similarities to our payments, indicating that the disaggregation by payment instruments can be valuable when using payment data for economic research. 

 \begin{figure}[h!]
    \centering
    \begin{subfigure}[b]{0.9\textwidth}
        \includegraphics[width=\textwidth]{inputs/shared_legend2_aggregate_timeseries.pdf}
    \end{subfigure}
    \begin{subfigure}[b]{0.32\textwidth}
        \centering
        \includegraphics[width=\textwidth]{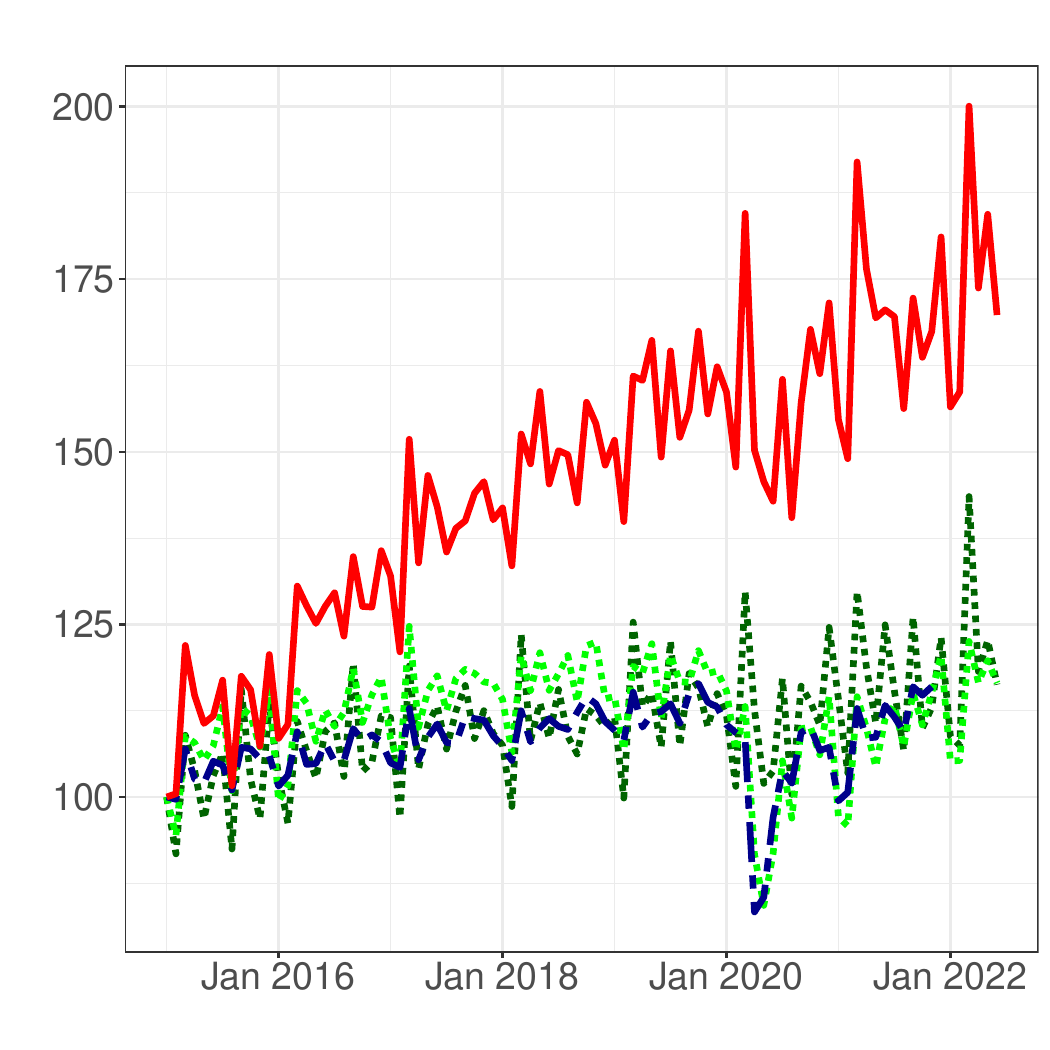}
        \caption{Value}
    \end{subfigure}
    \begin{subfigure}[b]{0.32\textwidth}
        \centering
        \includegraphics[width=\textwidth]{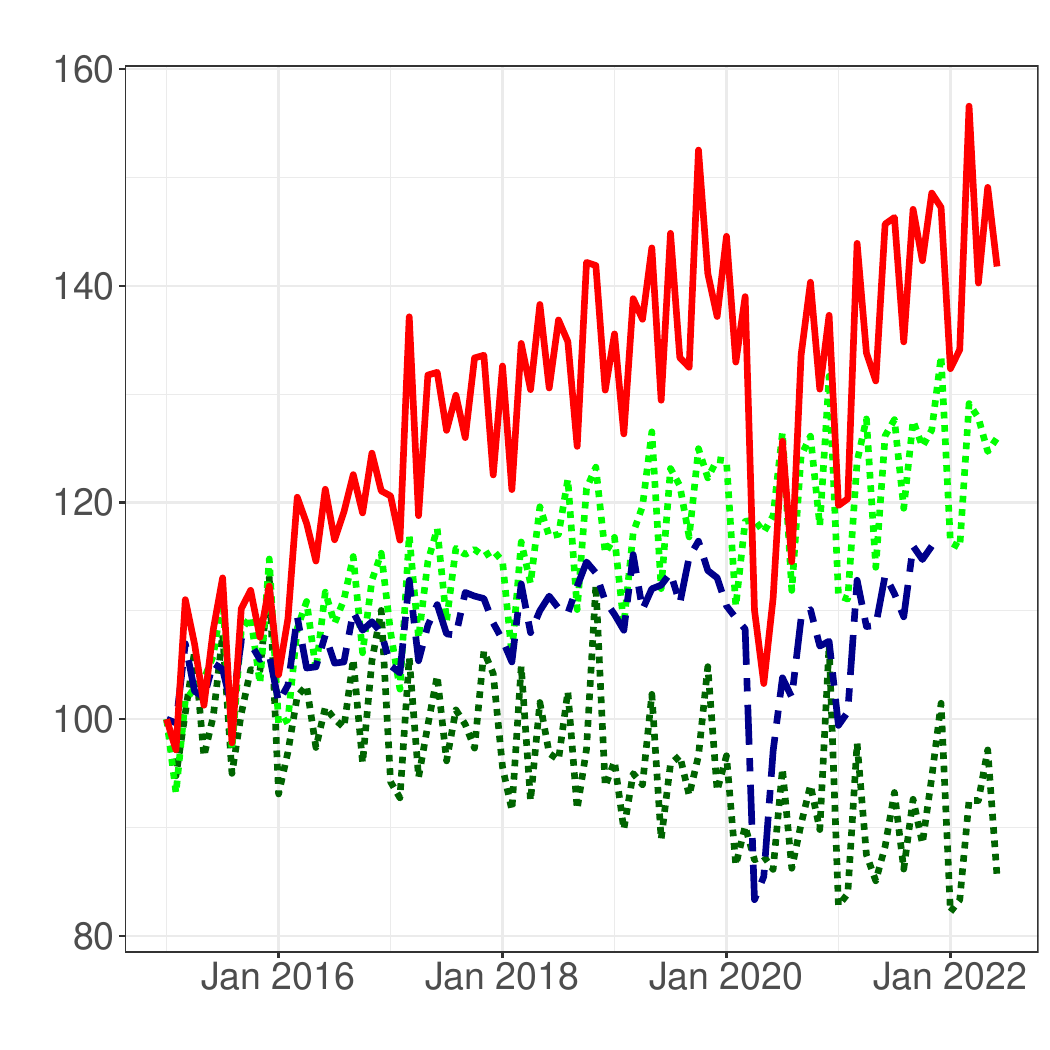}
        \caption{Counts}
    \end{subfigure}
    \begin{subfigure}[b]{0.32\textwidth}
        \centering
        \includegraphics[width=\textwidth]{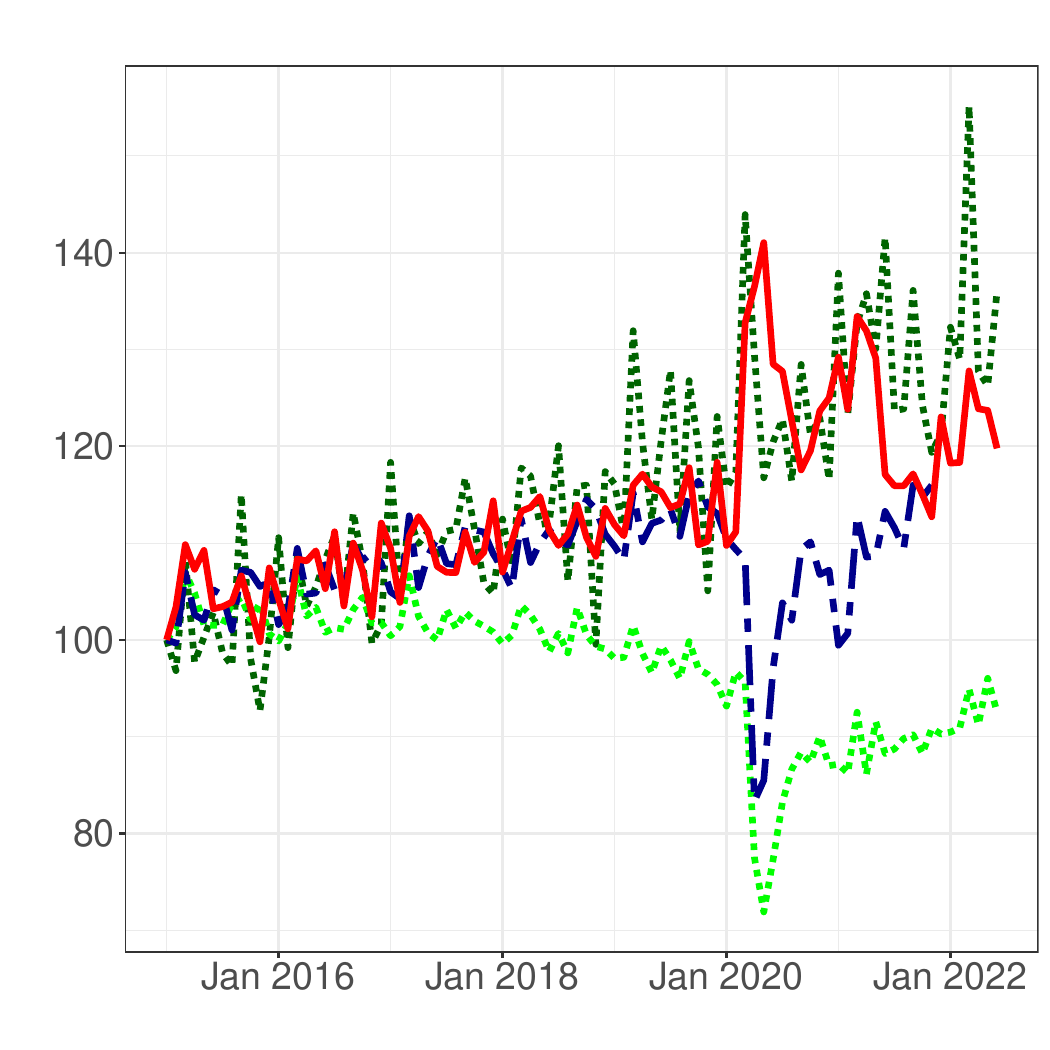}
        \caption{Average value}
    \end{subfigure}

    \caption{Monthly payments, Direct Debits and Credits}
    \label{app:fig:timeseries_aggr_index_payments_vs_bacs_instrumetns}

    \justifying \scriptsize
    \noindent
    Notes: These figures show monthly time series of the payment data, and Bacs transactions disaggregated by Direct Debits and Credits, indexed by 2015 = 100. The average value is calculated by dividing values by counts. The dark blue dashed line shows monthly deseasonalised GVA as a benchmark.

\end{figure}

Fig. \ref{app:fig:timeseries_aggr_index_payments_vs_bacs_instrumetns} shows a disambiguation of Bacs DD and DC in comparison to the trends in the payment data. While the Bacs aggregates grow only moderately by value at a similar pace as GDP, despite GDP being in real and Bacs in nominal terms, the payment data show a much steeper increase. While DC and DD evolve similarly by value, they differ by counts and average value. The number of Bacs DC decreased over time, but their average transaction value increased similarly than the average value in the payment data. This may be indicative of shifts in the utilisation of the Bacs system.\footnote{Possible reasons for changes in Bacs utilisation are the rise of FPS, increasing use of Cards, and digital payment innovation \citep{UKfinance2022UKpayments}, but answering this question is beyond the scope of this paper.}

\FloatBarrier
\subsection{Macroeconomic benchmarking}
\label{app:sec:macro}
Table \ref{app:tab:macro_benchmarking_inclCovid} shows the results of the macro-level correlation analysis introduced in Sec. \ref{sec:macro_benchmarking} when the period of Covid-19 is not removed from the data. While all correlations are lower, we observe still high values for the count data for almost all indicators except M1 and M3, suggesting that payment counts are more robust in capturing the dynamics of real economic indicators during the exceptional period of Covid-19. 

\begin{table}[h] \centering 
  \caption{Correlations with other payments and macro aggregates including Covid-19} 
  \label{app:tab:macro_benchmarking_inclCovid} 
\scriptsize 
\begin{tabular}{@{\extracolsep{5pt}} ccccccccc} 
\\[-1.8ex]\hline 
\hline \\[-1.8ex] 
 & Bacs & FPS & CHAPS & GDP nsa & GDP sa & M1 nsa & M3 nsa & Prices \\ 
\hline \\[-1.8ex] 

Share in 2019 & $0.207$ & $0.540$ & $0.013$ & $0.469$ & $0.469$ & $0.578$ & $0.363$ &\\ 
Share in 2021 & $0.221$ & $0.431$ & $0.013$ & $0.490$ & $0.514$ & $0.471$ & $0.321$ &\\ 

\hline \\[-1.8ex] 
\multicolumn{8}{l}{\emph{Raw data in levels}}\\
\hline \\[-1.8ex] 
Yearly (value) & $0.885$ & $0.964$ & $0.797$ & $0.159$ & $-0.380$ & $0.887$ & $0.916$ & $0.988$ \\ 
Monthly (value) & $0.824$ & $0.907$ & $0.696$ & $0.394$ & $0.484$ & $0.832$ & $0.868$ & $0.816$ \\ 
Yearly (count) & $0.941$ & $0.842$ & $0.948$ & $0.733$ & $-0.168$ & $0.629$ & $0.697$ & $0.756$ \\ 
Monthly (count) & $0.768$ & $0.739$ & $0.928$ & $0.799$ & $0.747$ & $0.636$ & $0.674$ & $0.645$ \\ 
Yearly (avg) & $0.476$ & $-0.525$ & $-0.024$ & $-0.472$ & $-0.507$ & $0.926$ & $0.904$ & $0.900$ \\ 
Monthly (avg) & $0.371$ & $-0.439$ & $0.098$ & $-0.326$ & $-0.072$ & $0.752$ & $0.773$ & $0.625$ \\ 
\hline \\[-1.8ex] 
\multicolumn{8}{l}{\emph{Growth rates}}\\
\hline \\[-1.8ex] 
Yearly (value) & $0.226$ & $-0.197$ & $0.288$ & $0.352$ & $0.160$ & $0.196$ & $0.354$ & $-0.526$ \\ 
Monthly (value) & $0.773$ & $0.613$ & $0.008$ & $0.709$ & $0.589$ & $-0.212$ & $-0.111$ & $0.010$ \\ 
Yearly (count) & $0.429$ & $-0.361$ & $0.066$ & $0.391$ & $0.207$ & $0.154$ & $0.324$ & $-0.480$ \\ 
Monthly (count) & $0.567$ & $0.526$ & $0.751$ & $0.893$ & $0.863$ & $-0.165$ & $-0.124$ & $0.199$ \\ 
Yearly (avg) & $-0.626$ & $-0.737$ & $0.582$ & $-0.947$ & $-0.656$ & $0.693$ & $0.533$ & $-0.771$ \\ 
Monthly (avg) & $-0.131$ & $-0.445$ & $0.590$ & $-0.813$ & $-0.823$ & $0.121$ & $0.141$ & $-0.343$ \\ 
\hline \\[-1.8ex] 

\end{tabular} 

\justifying \noindent \scriptsize
Notes: 
This table shows Pearson correlations between annual (monthly) payments and other UK payment schemes and macroeconomic aggregates (GDP, M1, M3, Prices) during 2016 and 2023 (08/2015 and 12/2023), including the Covid-19 period. ``sa'' (``nsa'') is short for (non-)seasonally adjusted. Our payment data and other payment aggregates are compared by aggregate values, counts, and average values (short ``avg'') given by value divided by count. 
Growth rates are calculated as percentage growth compared to the (same month of the) previous year (for monthly data). Bacs, FPS, and CHAPS data are obtained from \citet{payuk2023historicaldata}. 
Monthly GDP is proxied by indicative (non-)seasonally adjusted monthly ``Total Gross Value Added'' index data published by the ONS \citep{ons2023indicativeGDPdata, ons2023indicativeGDPadjusted}. 
``Prices'' is short for Consumer prices index data obtained from the OECD Key Economic Indicators (KEI) dataset \citep{oecd2023KEIdata}. M1 (M3) are narrow (broad) monetary aggregates, and thus nominal indicators, obtained from the OECD Main Economic Indicators (MEI) dataset \citep{oecd2023MEIdata}.
\end{table}

\FloatBarrier
\subsection{Aggregate network statistics}
\label{app:subsec:aggregate_network}

The Tables \ref{app:tab:network_stats_2019_threshold_pct_1} and \ref{app:tab:network_stats_2019_threshold_pct_5} summarise network statistics analogous to those in Table \ref{tab:network_stats_2019_no_truncation} for networks truncated at a 1\% and 5\% threshold level. The figure and the tables include the results for both truncation by input and output share. 
The density in all networks decreases in all networks, but with a much steeper slope for the ONS IOTs, which is due to the forestalled truncation caused by the SDC. The decrease is slightly faster in the output network, suggesting a higher concentration of outputs. 
The truncation also affects other properties of the networks, but qualitatively homogeneously across the different IOTs, except for assortativity. 

\begin{table}[!htbp] \centering 
  \caption{Properties of the payment and ONS-based IOTs in 2019, truncated with a 1\% threshold} 
  \label{app:tab:network_stats_2019_threshold_pct_1} 
\scriptsize 
\begin{tabular}{@{\extracolsep{1pt}} cccccc} 
\\[-1.8ex]\hline 
\hline \\[-1.8ex] 
Variable & Value & Count & PxP & SUT & IxI \\ 
\hline \\[-2ex]
\underline{\emph{Raw transactions -- truncation by input share} }\\  
Density & $0.157$ & $0.198$ & $0.181$ & $0.164$ & $0.186$ \\ 
Average degree & $16.214$ & $20.417$ & $17.546$ & $15.897$ & $18.072$ \\ 
Average strength & $5,808.344$ & $425,637.300$ & $8,567.096$ & $10,369.400$ & $8,344.459$ \\ 
Average weight & $358.239$ & $20,846.710$ & $488.254$ & $652.291$ & $461.730$ \\ 
Reciprocity & $0.212$ & $0.185$ & $0.224$ & $0.217$ & $0.235$ \\ 
Transitivity & $0.428$ & $0.488$ & $0.466$ & $0.433$ & $0.471$ \\ 
Assortativity by degree & $-0.310$ & $-0.429$ & $-0.266$ & $-0.216$ & $-0.248$ \\ 

\hline \\[-2ex]
\underline{\emph{Raw transactions -- truncation by output share} }\\  
Density & $0.115$ & $0.131$ & $0.129$ & $0.112$ & $0.139$ \\ 
Average degree & $11.796$ & $13.476$ & $12.536$ & $10.887$ & $13.505$ \\ 
Average strength & $5,963.682$ & $426,890.600$ & $8,009.429$ & $9,876.381$ & $7,888.582$ \\ 
Average weight & $505.563$ & $31,678.480$ & $638.910$ & $907.205$ & $584.116$ \\ 
Reciprocity & $0.170$ & $0.144$ & $0.214$ & $0.169$ & $0.232$ \\ 
Transitivity & $0.367$ & $0.394$ & $0.432$ & $0.393$ & $0.441$ \\ 
Assortativity by degree & $-0.308$ & $-0.364$ & $-0.096$ & $-0.054$ & $-0.154$ \\ 

\hline \\[-2ex]
\underline{\emph{Input shares -- truncation by input share} }\\  
Average strength & $0.805$ & $0.815$ & $0.740$ & $0.643$ & $0.708$ \\ 
Average weight & $0.050$ & $0.040$ & $0.042$ & $0.040$ & $0.039$ \\ 
\hline \\[-2ex]
\underline{\emph{Input shares -- truncation by output share} }\\ 
Average strength & $0.261$ & $0.204$ & $0.422$ & $0.427$ & $0.405$ \\ 
Average weight & $0.022$ & $0.015$ & $0.034$ & $0.039$ & $0.030$ \\ 
\hline \\[-2ex]
\underline{\emph{Output shares -- truncation by input share} }\\ 
Average strength & $0.351$ & $0.270$ & $0.530$ & $0.529$ & $0.507$ \\ 
Average weight & $0.022$ & $0.013$ & $0.030$ & $0.033$ & $0.028$ \\ 
\hline \\[-2ex]
\underline{\emph{Input shares -- truncation by output share} }\\ 
Average strength & $0.830$ & $0.814$ & $0.705$ & $0.642$ & $0.678$ \\ 
Average weight & $0.070$ & $0.060$ & $0.056$ & $0.059$ & $0.050$ \\ 
\hline \\[-1.8ex] 
\end{tabular}

\justifying \scriptsize \noindent
Notes: This table shows aggregate network statistics for networks truncated at a 1\% threshold, where links between industries are removed if the weight measured by the input (output) share is below 1\%. 
The first (second) column uses payment values (counts) as weights. The other columns represent official IOTs published by the ONS, where PxP is short for Product-by-Product, IxI for Industry-by-Industry, and SUT for Supply-and-Use Table. The data are aggregated into 105 distinct CPA codes (see Sec. \ref{subsec:harmonization_with_ONS}).
Raw transaction data are shown in £ million.

\end{table}

\begin{table}[!htbp] \centering 
  \caption{Properties of the payment and ONS-based IOTs in 2019, truncated with a 5\% threshold} 
  \label{app:tab:network_stats_2019_threshold_pct_5} 
\scriptsize 
\begin{tabular}{@{\extracolsep{1pt}} lccccc} 
\\[-1.8ex] \hline 
Variable & Value & Count & PxP & SUT & IxI \\ 
\hline \\[-2ex]
\underline{\emph{Raw transactions -- truncation by input share} }\\  
Density & $0.040$ & $0.045$ & $0.041$ & $0.034$ & $0.037$ \\ 
Average degree & $4.107$ & $4.650$ & $3.969$ & $3.299$ & $3.598$ \\ 
Average strength & $3,053.352$ & $236,213.000$ & $4,356.415$ & $5,378.907$ & $3,862.778$ \\ 
Average weight & $743.488$ & $50,793.190$ & $1,097.590$ & $1,630.481$ & $1,073.609$ \\ 
Reciprocity & $0.043$ & $0.063$ & $0.036$ & $0.075$ & $0.046$ \\ 
Transitivity & $0.136$ & $0.131$ & $0.156$ & $0.168$ & $0.133$ \\ 
Assortativity by degree & $0.400$ & $0.634$ & $0.245$ & $0.140$ & $0.248$ \\ 
\hline \\[-2ex]  \underline{\emph{Raw transactions -- truncation by output share}} \\ 
Density & $0.040$ & $0.033$ & $0.033$ & $0.031$ & $0.032$ \\ 
Average degree & $4.155$ & $3.447$ & $3.186$ & $2.990$ & $3.124$ \\ 
Average strength & $4,055.909$ & $278,181.800$ & $4,491.696$ & $5,933.608$ & $4,277.842$ \\ 
Average weight & $976.072$ & $80,711.900$ & $1,410.014$ & $1,984.690$ & $1,369.474$ \\ 
Reciprocity & $0.047$ & $0.028$ & $0.039$ & $0.041$ & $0.040$ \\ 
Transitivity & $0.134$ & $0.114$ & $0.137$ & $0.118$ & $0.118$ \\ 
Assortativity by degree & $0.470$ & $0.541$ & $0.246$ & $0.306$ & $0.304$ \\ 
\hline  \\[-2ex]  \underline{\emph{Input shares -- truncation by input share}} \\ 
Average strength & $0.530$ & $0.465$ & $0.437$ & $0.366$ & $0.391$ \\ 
Average weight & $0.129$ & $0.100$ & $0.110$ & $0.111$ & $0.109$ \\ 
\hline \\[-2ex]  \underline{\emph{Input shares -- truncation by output share}} \\  
Average strength & $0.086$ & $0.048$ & $0.169$ & $0.210$ & $0.164$ \\ 
Average weight & $0.021$ & $0.014$ & $0.053$ & $0.070$ & $0.052$ \\ 
\hline  \\[-2ex]  \underline{\emph{Output shares -- truncation by input share}} \\   
Average strength & $0.086$ & $0.065$ & $0.217$ & $0.238$ & $0.187$ \\ 
Average weight & $0.021$ & $0.014$ & $0.055$ & $0.072$ & $0.052$ \\  
\hline  \\[-2ex]  \underline{\emph{Output shares -- truncation by output share}} \\ 
Average strength & $0.660$ & $0.595$ & $0.505$ & $0.468$ & $0.456$ \\ 
Average weight & $0.159$ & $0.173$ & $0.158$ & $0.157$ & $0.146$ \\ 
\hline \\[-1.8ex] 
\end{tabular}

\justifying \scriptsize \noindent
Notes: This table shows aggregate network statistics for networks truncated at a 5\% threshold, where links between industries are removed if the weight measured by the input (output) share is below 5\%. 
The first (second) column uses payment values (counts) as weights. The other columns represent official IOTs published by the ONS, where PxP is short for Product-by-Product, IxI for Industry-by-Industry, and SUT for Supply-and-Use Table. The data are aggregated into 105 distinct CPA codes (see Sec. \ref{subsec:harmonization_with_ONS}).
Raw transaction data are shown in £ million. 
\end{table}

\FloatBarrier
\subsection{Auto- and cross-correlations}
\label{app:subsec:auto_cross_correlations}

\begin{figure}
    \centering

            \caption{Auto- and cross-correlations of inputs \& outputs}
        \label{app:fig:correlations_industry_level_input_output_raw}

        \includegraphics[width=\textwidth]{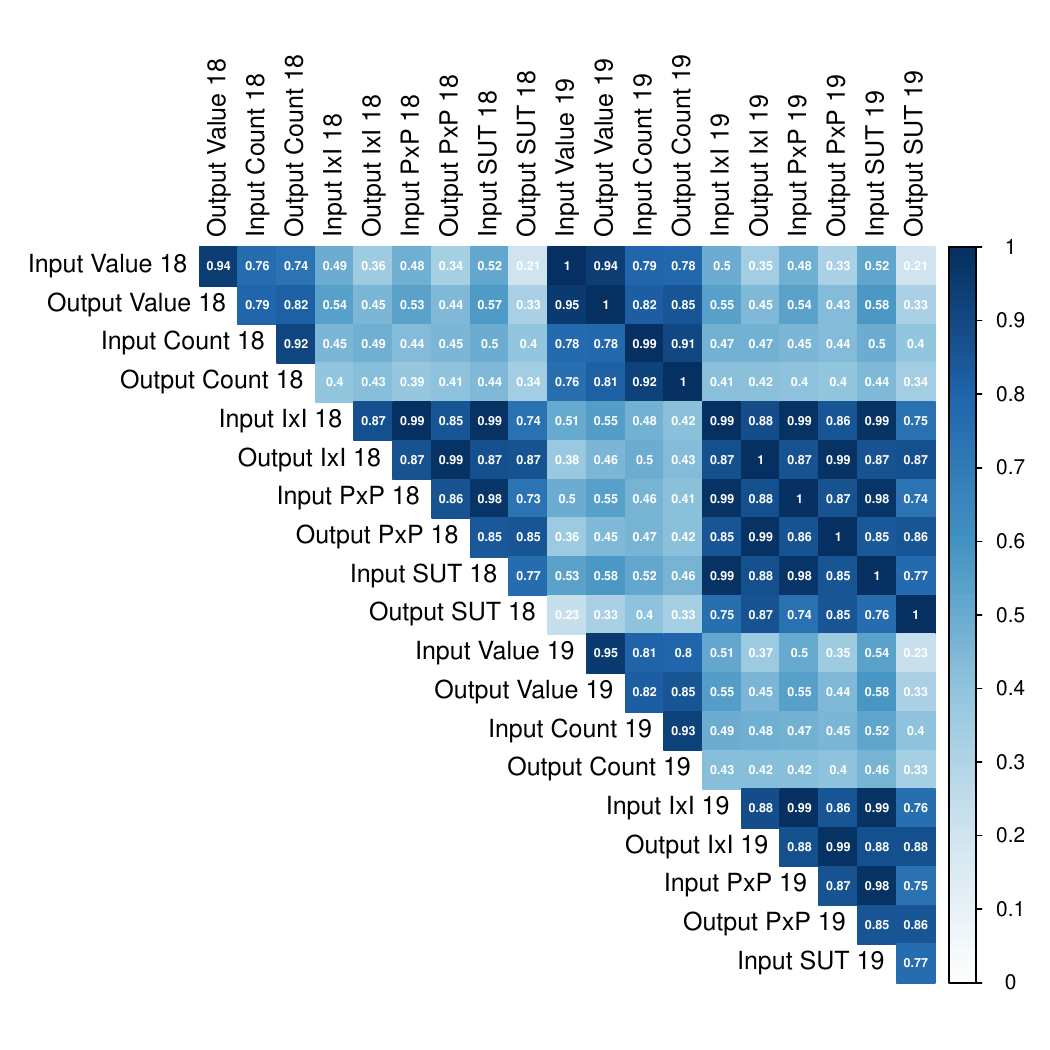}

\justifying \scriptsize \noindent

    Notes: The correlations are measured by the Pearson correlation coefficient between industry-level annual outputs and inputs in 2018-19 calculated by using raw transaction values and counts of the payment data and the row- and column sums of ONS IxI, PxP, and SUTs. 
\end{figure}

\begin{figure}

        \caption{Auto- and cross-correlations of input \& output growth}
        \label{app:fig:correlations_industry_level_input_output_raw_growth}

    \centering
        \includegraphics[width=\textwidth]{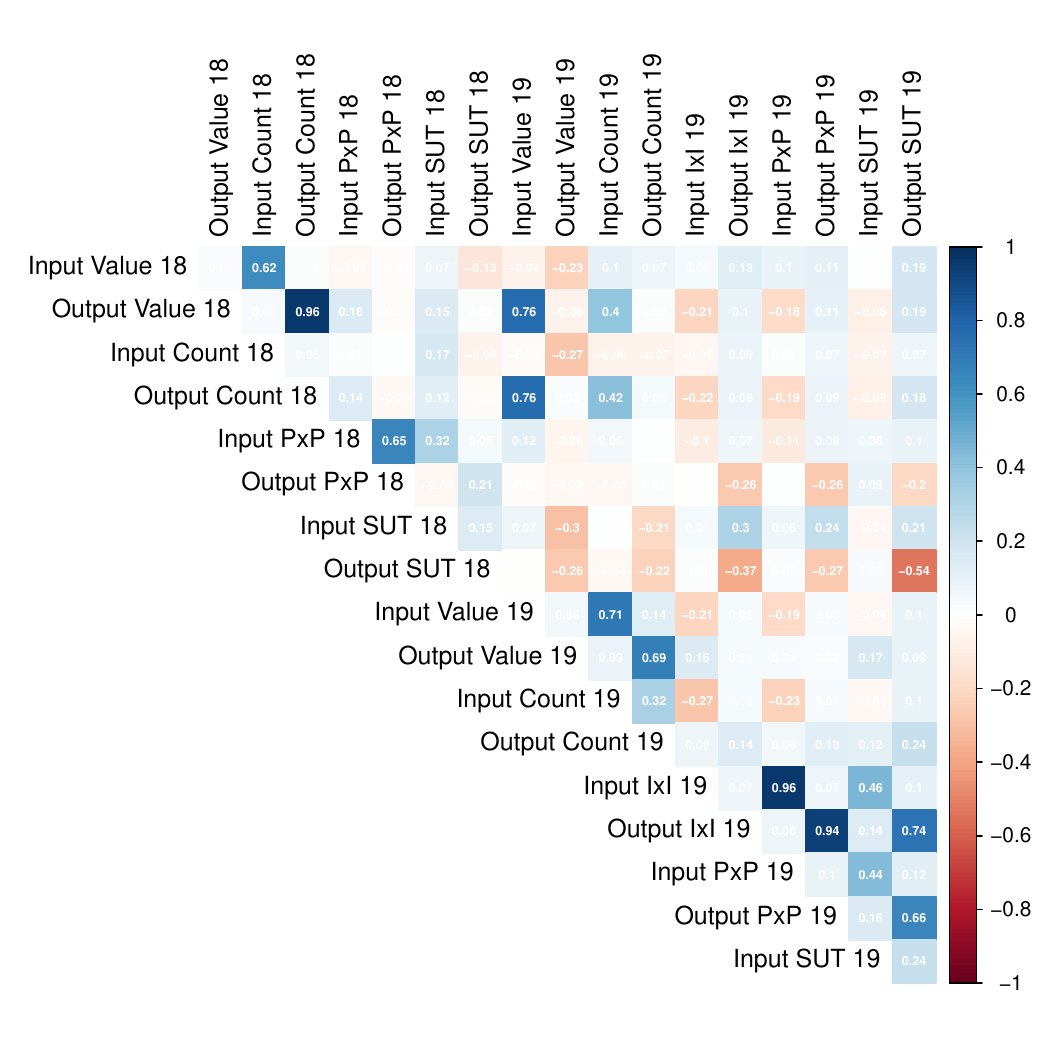}

\justifying \scriptsize \noindent
    Notes: The correlations are measured by the Pearson correlation coefficient between industry-level annual growth rates of outputs and inputs in 2018 and 2019 calculated by using raw transaction values and counts of the payment data and the row- and column sums of ONS IxI, PxP, and SUTs. 
\end{figure}

Fig. \ref{app:fig:correlations_industry_level_input_output_raw} shows the pairwise Pearson correlation coefficients for industry-level inputs and outputs derived from the Payment and ONS tables from 2018 and 2019, which can be seen as an indicator of industry size. As above, a darker colour indicates stronger correlations. An analogous figure for input and output growth rates is shown in Fig. \ref{app:fig:correlations_industry_level_input_output_raw_growth}. 

The analyses of growth rate correlations broadly confirm these relationships but with much lower correlation rates and discrepancies between in- and output growth, with output growth being much less or even negatively auto-correlated. Note that these correlations do not provide any information about statistical significance.  

At the industry level, we also correlated aggregate inputs and outputs in the payment data with other economic performance indicators, such as labour compensation and value added. As a broad takeaway, these analyses confirm that the payment data shows strong statistical relationships with these indicators. The correlations are weaker than those of ONS analogues, but a promising statistical signal remains confirming the data's value for economic analyses.

\FloatBarrier
\subsection{Scale differences across datasets}
\label{app:scale_differences}

\begin{figure}[h]
    \centering
    \includegraphics[width=0.8\textwidth]{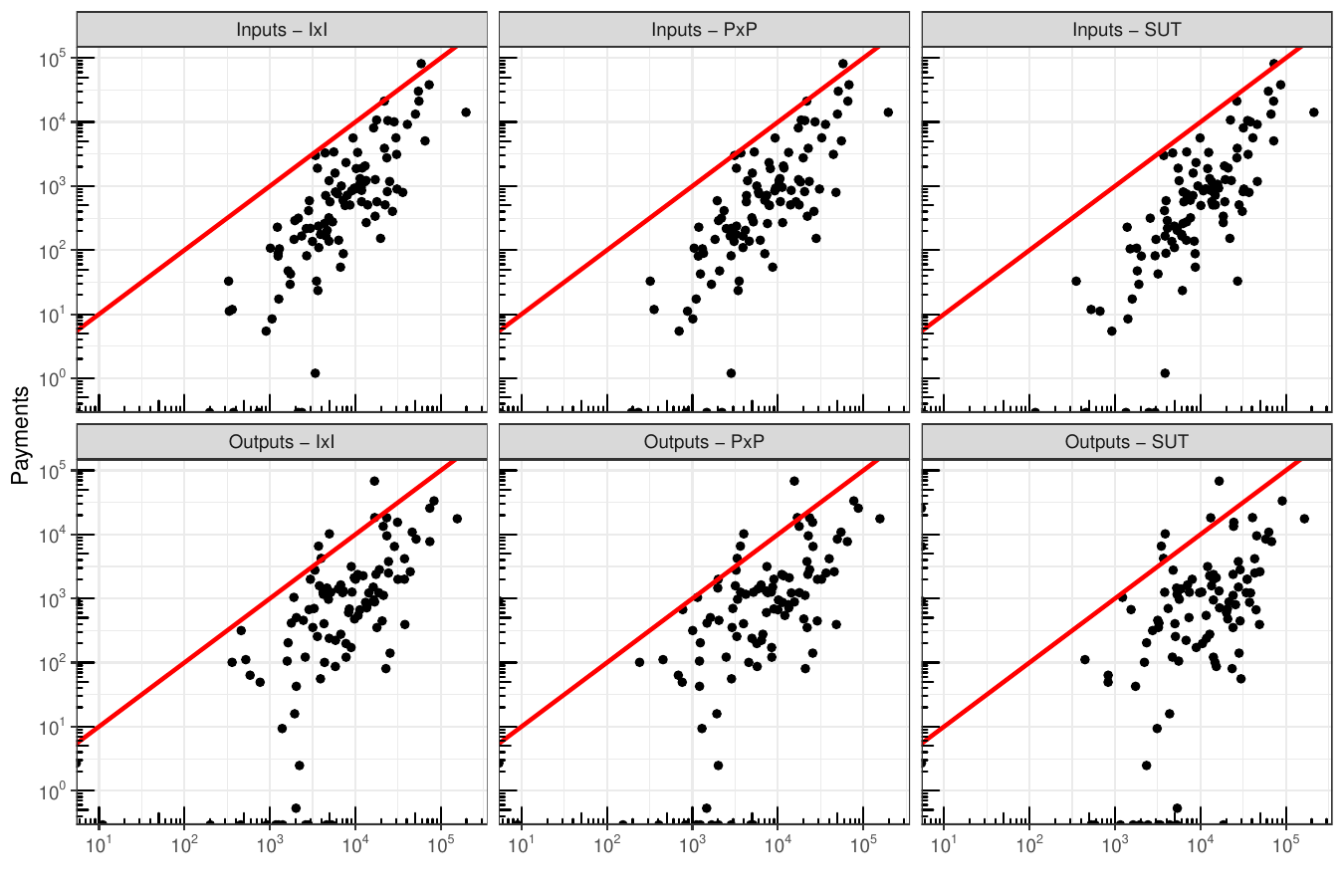}
    \caption{Comparison of industry sizes}
    \label{app:fig:input_output_scatterplot_2019}

   \justifying \scriptsize
    \noindent
    Notes: This figure shows the differences of industry-level aggregate inputs and outputs. Payment data values are shown at the vertical axis, and those for different ONS IOTs (IxI, PxP, SUT) at the horizontal. The red line is a 45 degree line, at which the values in the payment data would be equal to those in the ONS table. 
\end{figure}

Fig. \ref{app:fig:input_output_scatterplot_2019} illustrates the differences in the scale of aggregate input purchases and output sales, as captured by the different data sources. Each dot in the figure reflects the data for one of the 105 different industries. The red 45-degree line illustrates how the dots would be allocated if transaction values were equal. The figure shows that for the majority of sectors, we find much lower values in the payment data compared to the ONS datasets, with few exceptions.

\FloatBarrier
\subsection{Difference quantification}
\label{app:subsec:diff_quantification}
In this section, we provide additional analyses and details related to the difference analysis performed in Sec.~\ref{subsec:difference_quantification}. 
The proportional difference between the payment-based and the ONS IOTs, illustrated in Fig. \ref{fig:rel_diff_histogram_positives} is defined as

\begin{align}
    \log_{10}\varepsilon^{\text{ONS}}_{ij} = \Bigg|\log_{10}\left(\frac{  Z_{ij}^{\text{Value}} }{\sum_{i,j}Z^{\text{Value}}_{ij}} \right)- \log_{10} \left(\frac{  Z_{ij}^{\text{ONS}} }{\sum_{i,j}Z^{\text{ONS}}_{ij}} \right) \Bigg|,
    \label{app:eq:rel_difference_logs}
\end{align}

where $\text{ONS} \in \{ \text{IxI}, \text{PxP}, \text{SUT} \}$ and $\text{Value}$ corresponds to the payment-based table in values. 

We consider pairwise transactions that are non-zero in both datasets. 
To adjust for major differences in the scale and coverage, we normalise the transaction values between a pair of industries $i$ and $j$ by the aggregate value of all transactions in the respective IOT, excluding those between industry pairs that have no linkages in the other dataset. 

As an additional measure of difference, we also compile a scaled percentage difference measure, that allows keeping those links, that are non-zero in at least one of the data sets, using the formula
\begin{align}
    \tilde{\varepsilon}^{\text{ONS}}_{ij} = \log_{10} \left( \Bigg| \frac{  Z_{ij}^{\text{Value}} }{\sum_{i,j}Z^{\text{Value}}_{ij}} - \frac{  Z_{ij}^{\text{ONS}} }{\sum_{i,j}Z^{\text{ONS}}_{ij}} \Bigg| \cdot \frac{\sum_{i,j}(Z^{\text{Value}}_{ij} + Z^{\text{ONS}}_{ij})}{2} \right), 
    \label{app:eq:rel_difference_percent}
\end{align}

which compiles the absolute value of the difference in transactions measured as a percentage of total transactions in the respective dataset. We scale it by the average of the total number of transactions and take the log to deal with the highly skewed nature of the data.

\begin{figure}[!h]
    \centering

    \caption{Scaled percentage difference}
    \label{app:fig:difference_hist_percentage}

    \begin{subfigure}[]{0.9\textwidth}
        \centering

    \subcaption{One-sided zero-links included}
        \includegraphics[width=\textwidth]{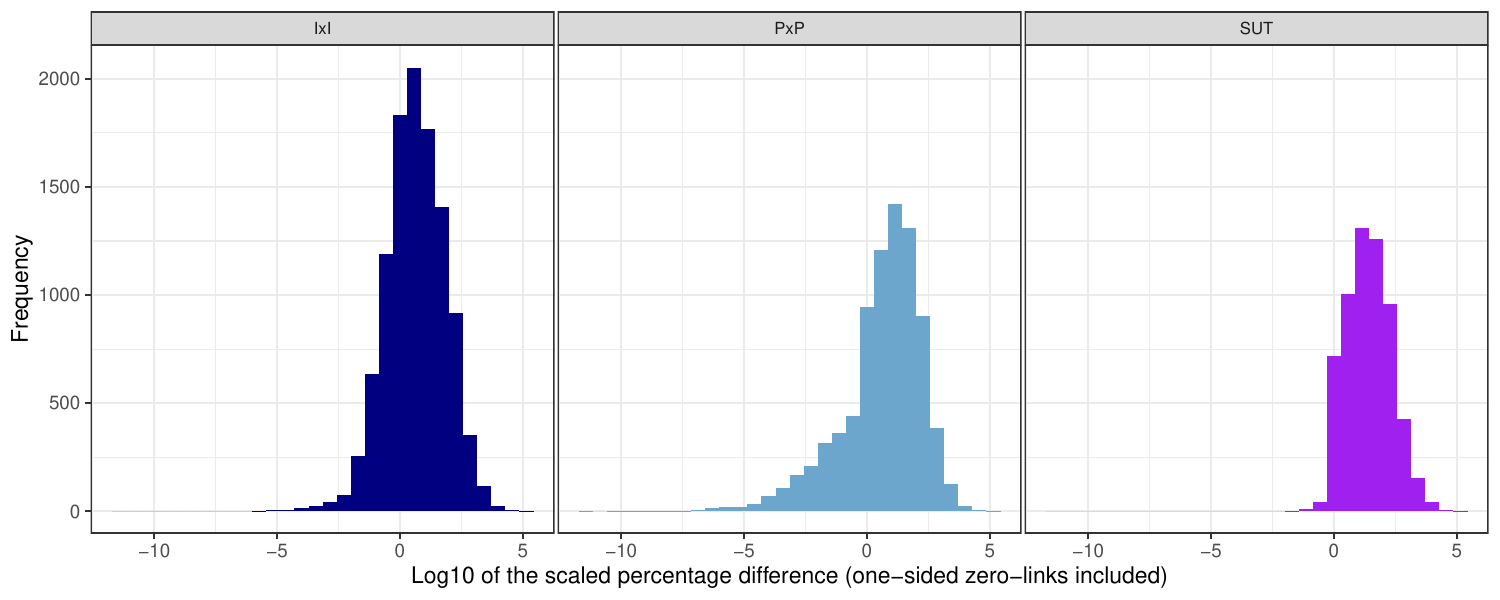}
    \end{subfigure}

    \begin{subfigure}[]{0.9\textwidth}
        \centering

    \subcaption{One-sided zero-links excluded}
        \includegraphics[width=\textwidth]{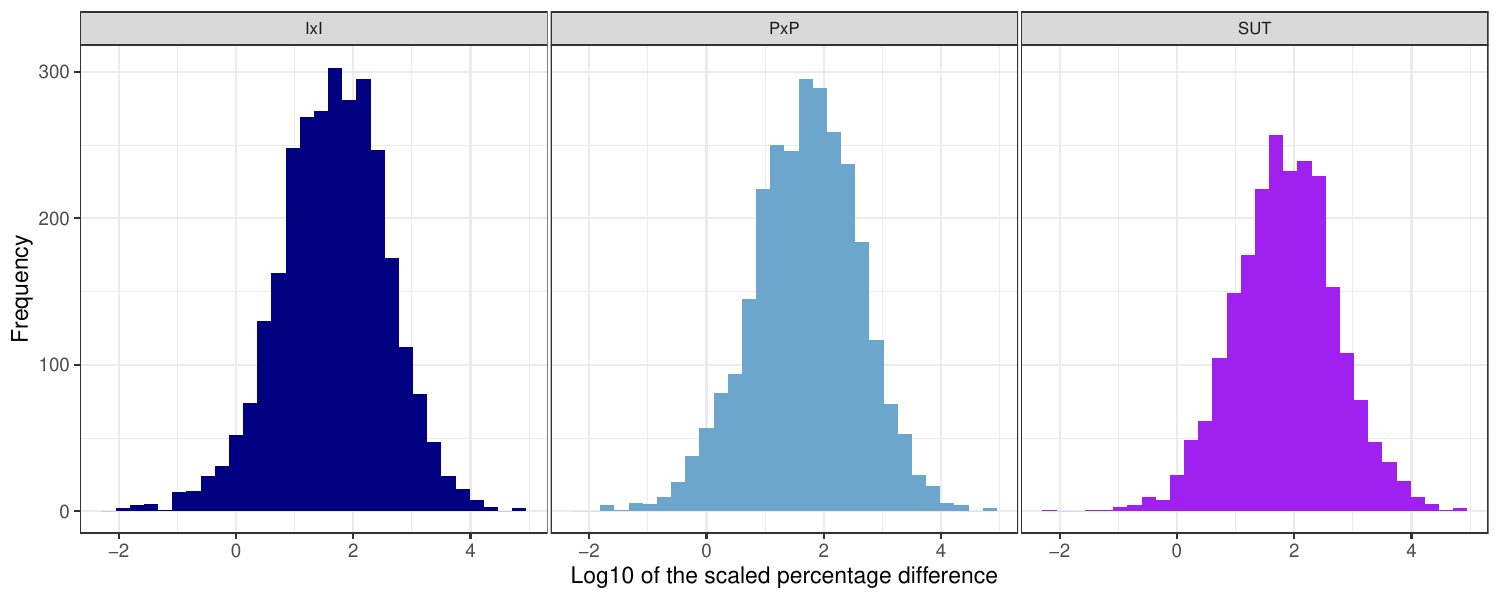}

    \end{subfigure}

    Notes: The figures show a comparison between the distribution of the scaled percentage edge-level differences (scaled at a log-10 basis) between the payment-based and ONS IOTs, when including or excluding links with a zero value in one of the two datasets. 
    
\end{figure}

\begin{table}[ht] \centering 
  \caption{Quartiles of the scaled percentage differences} 
  \label{app:tab:differences_quartiles_zerolinks} 
\begin{tabular}{ lp{1.5cm}p{1.5cm}p{1.5cm}p{1.5cm}} 
\\[-1.8ex]\hline 
\hline \\[-1.8ex] 
  & 25\% & 50\% & 75\% & 100\% \\ 
\hline \\[-1.8ex] 
IxI & $11.30$ & $47.81$ & $192.86$ & $62,674.01$ \\ 
including zero & $0.72$ & $4.34$ & $30.52$ & $73,521.71$ \\ 
\hline
PxP & $12.31$ & $53.50$ & $213.27$ & $63,817.68$ \\ 
including zero & $0.72$ & $8.05$ & $54.46$ & $73,543.50$ \\ 
\hline
SUT & $18.85$ & $72.75$ & $268.92$ & $81,455.65$ \\ 
including zero & $5.36$ & $24.42$ & $113.21$ & $88,906.08$ \\ 
\hline \hline \\[-1.8ex] 
\end{tabular} 

   \justifying \scriptsize
    \noindent
     Notes: Quantiles of the scaled percentage differences between the IxI, PxP, SUTs and the payment-based IOT in 2019. Unlike as in Fig. \ref{app:fig:difference_hist_percentage}, the values are not log-scaled. The scale of the scaled percentage difference is not comparable to the proportional difference used in the main text. 
\end{table} 

This modified measure is illustrated in Fig. \ref{app:fig:difference_hist_percentage} and summarised by quartiles in Table \ref{app:tab:differences_quartiles_zerolinks}. We find the differences to be much smaller when keeping the one-sided zero links in the data. 
Note that the scale of the indicator is not comparable to the difference metric used in the main text (Sec. \ref{subsec:difference_quantification}). In contrast to the proportional difference, the scaled percentage differences lack a clear quantitative interpretation but are used to illustrate the qualitative impact of removing one-sided zero-links. 

\FloatBarrier

\section{Additional material: stylised facts of the granular network}
\subsection{Correlation of growth rates}
\label{app:sec:corr_growth_rates}
In Sec.~\ref{sec:corr_growth_rates_main} we have studied the correlation of growth rates between different SIC-5 industries in our payment dataset and their dependence on network distance. To perform that analysis, we have truncated our network by imposing a threshold on the input shares.\footnote{A qualitatively similar result is also found when imposing a threshold on the output shares.}  
We truncate the network using an industry-specific approach by removing links that fall below a certain input share threshold, similar to above (see Fig. \ref{fig:network_density_plot_3digit} in the main text). In detail, this implies:
\begin{itemize}
    \item Aggregate all monthly transactions in a given year at a given level of industry aggregation to obtain a network of yearly transactions.
    \item Impose a threshold below which to remove links. The threshold is specified through the input shares of a given industry, in line with the prescription used by \cite{carvalho2014from} for a similar analysis.
\end{itemize}
We do this for each year from 2016 to 2023 and use the truncated annual network to calculate the distances. We transform the network into an edgelist and match the annual growth rate of the selling and buying industry to the respective pairwise distance in the network from the same year. For example, growth rates from 2019 are attributed to the distances calculated from the network in 2019. 

\FloatBarrier

\subsection{Influence vector and power law}
\label{app:subsec:influence_power_law}
Table \ref{app:tab_plfit_stats} summarises the test statistics and fitted coefficients, when fitting a power law function to the influence vector for both the payment network of values and counts, and for the years illustrated in Fig. \ref{fig:ccdf_influence_vector}. The bottom panel of the table shows the result for truncated data. We observe $\gamma$-values ranging between 1.34 and 1.69 (0.98 and 2.21) for the network of payments in values (counts). Generally, we find that the power law hypothesis is not significant for the value data but holds for some years when using the count data. Truncating the data does not have any relevant effect. Also, we made additional robustness checks considering all available years and compiled the influence vector using slightly different but plausible assumptions of the labour share $\alpha_L = \{ 0.3, 0.7 \}$, and did not find any qualitative change compared to the results shown here.  

\begin{table}[!htbp] \centering 
\footnotesize
  \caption{Power law fitting statistics} 
  \label{app:tab_plfit_stats} 
\begin{tabular}{@{\extracolsep{5pt}} c|ccccc|ccccc} 
\multicolumn{9}{l}{}\\[-1.8ex]\hline 
\hline \multicolumn{9}{l}{}\\[-1.8ex] 
\multicolumn{1}{c|}{}&\multicolumn{4}{c}{Value}&\multicolumn{4}{c}{Count}\\
\hline \\[-1.8ex] 
Year & $\gamma$ & xmin& logLik& KS.stat& p-value&$\gamma$& xmin& logLik& KS.stat& p-value\\ 
\hline \\[-1.8ex] 
2017 & 1.362 & 0.001 & 530.532 & 0.06 & 0.855 & 2.082 & 0.001 & 1570.413 & 0.167 & 0 \\ 
2019 & 1.429 & 0.001 & 713.972 & 0.087 & 0.267 & 1.141 & 0.003 & 133.594 & 0.072 & 0.995 \\ 
2021 & 1.615 & 0.001 & 1057.93 & 0.088 & 0.117 & 1.974 & 0.001 & 1539.515 & 0.17 & 0 \\ 
2023 & 1.382 & 0.001 & 767.922 & 0.111 & 0.061 & 0.982 & 0.001 & 295.351 & 0.062 & 0.964 \\ 
\hline \\[-1.8ex] 
&\multicolumn{8}{c}{Data truncated at 10\% quantile of transaction value}  \\ 
\hline \\[-1.8ex] 
2017 & 1.343 & 0.001 & 474.976 & 0.054 & 0.952 & 2.207 & 0.001 & 1646.743 & 0.172 & 0 \\ 
2019 & 1.455 & 0.001 & 677.768 & 0.086 & 0.305 & 1.882 & 0.001 & 1345.955 & 0.167 & 0 \\ 
2021 & 1.689 & 0.001 & 1098.346 & 0.086 & 0.117 & 1.022 & 0.001 & 319.632 & 0.062 & 0.95 \\ 
2023 & 1.49 & 0.001 & 857.5 & 0.117 & 0.03 & 1.022 & 0.001 & 309.591 & 0.058 & 0.977 \\ 
\hline \multicolumn{9}{l}{}\\[-1.8ex] 
\end{tabular}

\justifying \noindent \scriptsize
Notes: 
This table shows the power law fitting statistics, where $\gamma$ is the fitted exponent, xmin is the minimum level of the influence vector beyond which a power law can be reasonably fitted (see \citet{Clauset_2009}), logLik shows the log-Likelihood, and KS is short for the Kolmogorov-Smirnov test statistic for significance. The p-value indicates the probability of rejecting the hypothesis that the distribution of the influence vector could have been drawn from a power law distribution. A p-value $<$0.05 supports the power law hypothesis.  

\end{table}

\begin{table}[!htbp] \centering 
\footnotesize
  \caption{Top 10 industries by influence vector} 
  \label{app:tab:top10_influence_vector} 
\begin{tabular}{@{\extracolsep{5pt}} ccl|ccl} 
\\[-1.8ex]\hline 
\hline \\[-1.8ex] 
SIC & & Industry description & SIC & & Industry description\\ 
\hline \\[-1.8ex] 
\multicolumn{3}{l|}{2017}&&&\\ 
\multicolumn{3}{c|}{Value}&\multicolumn{3}{c}{Count}\\ 
\hline \\[-1.8ex] 
84110 & 0.1273 & General public administration & 84110 & 0.0719 & General public administration \\ 
82990 & 0.0538 & Other business support services n.e.c. & 82990 & 0.0574 & Other business support services n.e.c. \\ 
64999 & 0.0347 & Financial intermediation n.e.c. & 64910 & 0.035 & Financial leasing \\ 
64910 & 0.0226 & Financial leasing & 61900 & 0.0347 & Other telecommunications \\ 
65110 & 0.0187 & Life insurance & 64999 & 0.0304 & Financial intermediation n.e.c. \\ 
61900 & 0.0181 & Other telecommunications & 45111 & 0.0223 & Sale of new \& motor vehicles \\ 
45111 & 0.0175 & Sale of new \& motor vehicles & 64191 & 0.0195 & Banks \\ 
70100 & 0.0095 & of head offices & 65110 & 0.0152 & Life insurance \\ 
62090 & 0.0087 & Other information technology services & 64921 & 0.0135 & Credit granting by non-deposit finance\\ 
49410 & 0.0074 & Freight transport by road & 62090 & 0.0121 & Other information technology services \\ 
\hline \\[-1.8ex] 
\hline \\[-1.8ex] 
\multicolumn{3}{l|}{2023}&&&\\ 
\multicolumn{3}{c|}{Value}&\multicolumn{3}{c}{Count}\\ 
\hline \\[-1.8ex] 
84110 & 0.1146 & General public administration & 84110 & 0.0946 & General public administration \\ 
82990 & 0.04 & Other business support services n.e.c. & 82990 & 0.0423 & Other business support services n.e.c. \\ 
65110 & 0.0315 & Life insurance & 64910 & 0.0346 & Financial leasing \\ 
64999 & 0.0287 & Financial intermediation n.e.c. & 61900 & 0.0323 & Other telecommunications \\ 
64910 & 0.0206 & Financial leasing & 45111 & 0.0254 & Sale of new \& motor vehicles \\ 
61900 & 0.018 & Other telecommunications & 64999 & 0.0207 & Financial intermediation n.e.c. \\ 
45111 & 0.0173 & Sale of new \& motor vehicles & 62090 & 0.0204 & Other information technology services \\ 
62090 & 0.0133 & Other information technology services & 65110 & 0.0133 & Life insurance \\ 
35130 & 0.0113 & Distribution of electricity & 64921 & 0.0125 & Credit granting by non-deposit finance \\ 
49410 & 0.0088 & Freight transport by road & 35130 & 0.0112 & Distribution of electricity \\ 
\hline \\[-1.8ex] 
\end{tabular} 
\end{table} 

Table \ref{app:tab:top10_influence_vector} also shows the industries that would be ranked as most central for 2017 and 2023. Consistent with earlier results and the conceptual discussion, we find the public sector, finance, trade and retail sectors to be highly central, which is different from centrality in IOTs following the NA standards. 

\end{document}